\documentclass[a4paper,onecolumn,accepted=2022-03-04]{quantumarticle}
\pdfoutput=1

\usepackage[utf8]{inputenc}
\usepackage[english]{babel}
\usepackage[T1]{fontenc}
\usepackage{amsmath,amssymb}
\usepackage{amsthm}
\usepackage{xspace}
\usepackage{qcircuit}
\usepackage{url} 
\usepackage{verbatim} 
\usepackage[linesnumbered ,boxed]{algorithm2e}

\usepackage{color,graphicx}

\usepackage{hyperref}
\usepackage{cleveref}

\begin{document}
\newcommand{\handout}[5]{
   \renewcommand{\thepage}{#1-\arabic{page}}
   \noindent
   \begin{center}
   \framebox{
      \vbox{
    \hbox to 5.78in { {\bf Advanced Algorithms - Lecture} \hfill #2 }
       \vspace{4mm}
       \hbox to 5.78in { {\Large \hfill #5  \hfill} }
       \vspace{2mm}
       \hbox to 5.78in { {\it #3 \hfill #4} }
      }
   }
   \end{center}
   \vspace*{4mm}
}

\newcommand{\lecture}[4]{\handout{#1}{#2}{Lecturer:
#3}{Scribe: #4}{Lecture #1}}
\newcommand{\recitation}[4]{\handout{#1}{#2}{Lecturer:
#3}{Scribe: #4}{Recitation #1}}
\newtheorem{theorem}{Theorem}
\newtheorem{corollary}[theorem]{Corollary}
\newtheorem{lemma}[theorem]{Lemma}
\newtheorem{observation}[theorem]{Observation}
\newtheorem{proposition}{Proposition}
\newtheorem{definition}[theorem]{Definition}
\newtheorem{claim}[theorem]{Claim}
\newtheorem{fact}[theorem]{Fact}
\newtheorem{assumption}[theorem]{Assumption}
\newtheorem{example}{Example}

\newenvironment{proof-sketch}{\noindent{\bf  Proof sketch:}\hspace*{1em}}{\qed\bigskip}
\newenvironment{proof-idea}{\noindent{\bf Proof Idea:}\hspace*{1em}}{\qed\bigskip}
\newenvironment{proof-of-lemma}[1]{\noindent{\bf Proof of Lemma #1:}\hspace*{1em}}{\qed\bigskip}
\newenvironment{proof-attempt}{\noindent{\bf Proof Attempt:}\hspace*{1em}}{\qed\bigskip}
\newenvironment{proofof}[1]{\noindent{\bf Proof
of #1:}\hspace*{1em}}{\qed\bigskip}
\newenvironment{remark}{\noindent{\bf Remark}\hspace*{1em}}{\bigskip}

\newcommand{\tr}[1]{Tr(#1)} 
\newcommand{\locHam}{\textsc{local hamiltonian}}
\newcommand{\YES}{yes}
\newcommand{\NO}{no}
\newcommand{\trace}{\text{tr}}

\newcommand{\id}{\mathbb{I}}

\newcommand{\bra}[1]{\langle #1|}
\newcommand{\ket}[1]{|#1\rangle}
\newcommand{\braket}[2]{\langle #1|#2\rangle}
\newcommand{\ketbra}[2]{\ket{#1}{\bra{#2}}}

\newcommand{\NN}{\mathbb N}

\renewcommand{\P}{\textsf{P}\xspace}
\newcommand{\NP}{\textsf{NP}\xspace}
\newcommand{\NPC}{\textsf{NP-Complete}\xspace}
\newcommand{\NPH}{\textsf{NP-Hard}\xspace}
\newcommand{\CH}{\textsf{C-Hard}\xspace}
\newcommand{\C}{\textsf{C}\xspace}
\newcommand{\D}{\textsf{D}\xspace}
\renewcommand{\DH}{\textsf{D-Hard}\xspace}
\newcommand{\UNP}{\textsf{UP}\xspace}
\newcommand{\UNPC}{\textsf{UP-Complete}\xspace}
\newcommand{\RP}{\textsf{RP}\xspace}
\newcommand{\BPP}{\textsf{BPP}\xspace}
\newcommand{\PP}{\textsf{PP}\xspace}
\newcommand{\Pc}{\textsf{P}\xspace}
\newcommand{\MA}{\textsf{MA}\xspace}
\newcommand{\MAC}{\textsf{MA-Complete}\xspace}
\newcommand{\UMA}{\textsf{UMA}\xspace}
\newcommand{\UMAC}{\textsf{UMA-Complete}\xspace}
\newcommand{\QMA}{\textsf{QMA}\xspace}
\newcommand{\FewQMA}{\textsf{Few-QMA}\xspace}
\newcommand{\PreciseQMA}{\textsf{PreciseQMA}\xspace}
\newcommand{\PrecisePGQMA}{\textsf{PrecisePGQMA}\xspace}

\newcommand{\QMAC}{\textsf{QMA-Complete}\xspace}
\newcommand{\UQMA}{\textsf{UQMA}\xspace}
\newcommand{\UQMAC}{\textsf{UQMA-Complete}\xspace}
\newcommand{\QCMA}{\textsf{QCMA}\xspace}
\newcommand{\QCMAC}{\textsf{QCMA-Complete}\xspace}
\newcommand{\UQCMA}{\textsf{UQCMA}\xspace}
\newcommand{\UQCMAC}{\textsf{UQCMA-Complete}\xspace}
\newcommand{\BQP}{\textsf{BQP}\xspace}
\newcommand{\coUP}{\textsf{co-UP}\xspace}
\newcommand{\PGQMA}{\textsf{PGQMA}\xspace}
\newcommand{\PGQMAC}{\textsf{PGQMA-Complete}\xspace}
\newcommand{\GQMA}{\textsf{GQMA}\xspace}
\newcommand{\GQCMA}{\textsf{GQCMA}\xspace}
\newcommand{\GMA}{\textsf{GMA}\xspace}
\newcommand{\SZK}{\textsf{SZK}\xspace}
\newcommand{\SZKC}{\textsf{SZK-Complete}\xspace}
\newcommand{\PSPACE}{\textsf{PSPACE}\xspace}

\newcommand{\FACTORING}{\textsf{factoring}}

\newcommand{\poly}{\textsf{poly}}

\newcommand{\lh}{\textsc{local hamiltonian}}
\newcommand{\onedlh}{\textsc{1-d hamiltonian}}
\newcommand{\ulh}{\textsc{unique local hamiltonian}}
\newcommand{\pglh}{\textsc{poly-gap local hamiltonian}}
\newcommand{\onedulh}{\textsc{unique 1-d hamiltonian}}
\newcommand{\onedcgulh}{\textsc{1-d constant-gap local hamiltonian}}
\newcommand{\onedpglh}{\textsc{1-d poly-gap local hamiltonian}}
\newcommand{\gappedlocHam}{\textsc{gapped local hamiltonian}}
\newcommand{\sat}{\textsc{sat}}
\newcommand{\spectralgap}{\textsc{spectral gap}}
\newcommand{\tnpp}{\textsc{tnpp}}
\newcommand{\unppp}{\textsc{unppp}}
\renewcommand{\L}{\textsc{l}}
\newcommand{\tmapp}{\textsc{tmapp}}
\newcommand{\umapp}{\textsc{umapp}}
\newcommand{\tqcmapp}{\textsc{tqcmapp}}
\newcommand{\uqcmapp}{\textsc{uqcmapp}}

\title{The Pursuit of Uniqueness: Extending Valiant-Vazirani Theorem to the Probabilistic and Quantum Settings}
\author{Dorit Aharonov} 
\affiliation{School of Computer Science and Engineering, 
The Hebrew University, Jerusalem, Israel.}  
\email{doria@cs.huji.ac.il}

\author{Michael Ben-Or}
\affiliation{School of Computer Science and Engineering, 
The Hebrew University, Jerusalem, Israel.} 
\email{benor@cs.huji.ac.il}

\author{Fernando G.S.L. Brand\~ao}
\affiliation{AWS Center for Quantum Computing, Pasadena, CA 91125, USA}
\affiliation{IQIM, California Institute of Technology, Pasadena, CA 91125, USA. }
\email{fbrandao@amazon.com}

\author{Or Sattath} 
\affiliation{Computer Science Department, 
Ben-Gurion University of the Negev, Beer-Sheva, Israel.}
\email{sattath@bgu.ac.il.}

\maketitle

\begin{abstract}
Valiant-Vazirani showed in 1985 \cite{VV85} that solving \NP\ with the promise that ``yes'' instances have only one witness is powerful enough to solve the entire \NP\ class (under randomized reductions). 

We are interested in extending this result to the quantum setting. 
We prove extensions to the classes Merlin-Arthur ($\MA$) and 
Quantum-Classical-Merlin-Arthur ($\QCMA$) \cite{AN02}. 
Our results have implications for the complexity of approximating the ground state energy of a quantum local Hamiltonian 
with a unique ground state and an \textit{inverse polynomial} spectral gap. We show that the estimation (to within polynomial accuracy) of the ground state energy of poly-gapped 1-D local Hamiltonians is $\QCMA$-hard, under randomized reductions. This is in stark contrast to the case of constant gapped 1-D Hamiltonians, which is in $\NP$ \cite{Has07}. 
Moreover, it shows that unless $\QCMA$ can be reduced to $\NP$ by randomized 
reductions, there is no classical description of the ground state of every 
poly-gapped local Hamiltonian that allows efficient calculation of expectation values. 

Finally, we discuss a few of the obstacles to the establishment of an analogous result to the class 
Quantum-Merlin-Arthur ($\QMA$). In particular, we show that random projections fail to provide a 
polynomial gap between two witnesses.
\end{abstract}

\section{Introduction}

\subsection{Extending Valiant-Vazirani}
One of the properties of the class \NP\ is that the number of witnesses 
may vary from zero to exponentially many.
How difficult is it to distinguish between ``no'' instances and ``yes''
instances that have a unique witness? Though one might think that such a problem is easier than solving \NP, the celebrated result of Valiant and Vazirani~\cite{VV85} proved that it is \emph{not} much easier. Their main result can be stated as follows:
\begin{theorem}[\cite{VV85}] 
\label{thm:valiant-vazirani}
There exists a $\UNP$ promise problem that is $\NPH$ under randomized reductions. 
\end{theorem}
In the above, $\UNP$ is the class containing all promise problems for which a ``yes'' instance has a unique accepting witness --- see Definition~\ref{def:unp}, the promise problem that is shown to exist is given in Definition~\ref{def:tnpp}, and randomized reductions are introduced formally in Definition~\ref{def:randomized_reduction}.

The classes $\MA$, \QCMA\ and \QMA\ are probabilistic and quantum analogues of \NP. Informally, we say a problem is in $\MA$ if for every ``yes'' instance there is a witness that the polynomial verifier accepts with high probability (e.g., in the range $(\frac{2}{3}, 1)$), while for ``no'' instances she only accepts with a small probability (e.g., in (0, 1/3)), no matter which witness is given to her. The class $\QCMA$ is similarly defined, but now the verifier can use a quantum computer to decide whether or not to accept the witness. In $\QMA$, not only does the verifier use a quantum computer to verify the proof, but also the proof itself is a quantum state comprising a polynomial (in the input size) number of qubits. The formal definitions are in Section~\ref{sec:preliminary_definitions}.

We define $\UMA$ and $\UQCMA$ as the restrictions of $\MA$ and $\QCMA$, respectively, to instances with a 
unique accepting witness. Roughly speaking, in a ``yes'' instance of 
a problem in $\UMA$ or $\UQCMA$, \textit{one} proof convinces the verifier with probability larger than $\frac{2}{3}$, while any other witness will cause the verifier to \emph{reject} with a probability of at least $\frac{2}{3}$. 
In a ``no'' instance, the verifier will reject all witnesses with a probability of at least $\frac{2}{3}$. We similarly  define the class \UQMA\, the unique variant of $\QMA$: the 
conditions for a ``no'' instance are the same as in $\QMA$, but for a ``yes''
instance, we demand that a state $\ket{\psi}$ exists and that it is accepted with a probability of at least $\frac{2}{3}$, while all states $\ket{\varphi}$ orthogonal to $\ket{\psi}$ are 
rejected with a probability of at least $\frac{2}{3}$.

We can ask a similar question to that posed by Valiant \& Vazirani 
about each of these classes: is it easier to solve $\UMA$ (respectively, $\UQCMA$ and $\UQMA$) promise problems than it is to solve $\MA$ (respectively, $\QCMA$ and $\QMA$)?
The quantum-related questions are also motivated by physical questions about the ground states
of local Hamiltonians. To that end, we provide some interesting
implications that we describe below. 

In this paper, we partially answer these questions by presenting generalizations of the Valiant-Vazirani theorem to $\MA$ and $\QCMA$. In addition, we discuss some of the challenges of achieving a similar result for $\QMA$, which remains an open problem.

\begin{theorem}
\label{thm:uma_problem_which_is_MA_hard}
There exists a promise problem (specified in Definition~\ref{def:tmapp}) in $\UMA$ that is $\MA$-hard under randomized reductions. 
\end{theorem}

\begin{theorem}
\label{thm:uqcma_problem_which_is_QCMA_hard}
There exists a promise problem (specified in Definition~\ref{def:tqcmapp}) in $\UQCMA$ that is $\QCMA$-hard under randomized reductions. 
\end{theorem}
The full proofs of Theorems~\ref{thm:uma_problem_which_is_MA_hard} and~\ref{thm:uqcma_problem_which_is_QCMA_hard} are given in Sections~\ref{sec:ma} and~\ref{sec:vv_to_qcma}, respectively. Both proofs rely heavily on the Valiant-Vazirani construction \cite{VV85} (see also~\cite[Section 17.4.1]{AB09} for another simple proof). We present a proof for the (original) Valiant-Vazirani theorem in three attempts, each of which improves on a shortcoming of its predecessor:
\begin{enumerate}
	\item The reduction guesses the size of the accepting witness set, and uses a random ``filter'' with a certain degree of screening that is determined by the set size. If the size of the accepting set is $w$, we add each potential witness to a random set with probability $1/w$, and everything outside that set is filtered out. If we correctly guess the size of the accepting witness set with a constant probability, exactly one valid witness will pass the filter. 
	\item We observe that it is not essential to guess the exact size of the witness set; a multiplicative approximation is adequate. Using this approach reduces the possible number of guesses from exponentially many in the previous attempt to linear (in the witness length).
	 \item We replace the random ``filter'' with a pseudorandom ``filter'' --- a pairwise-independent hash function --- without losing any of the properties. Moreover, a pairwise-independent hash-function has a polynomial size description and can be computed efficiently (unlike truly random subsets of $\{0,1\}^n$). This is crucial so that the reduction runs in (randomized) polynomial time and that the verifier is efficient.
\end{enumerate}
These arguments are made precise in Section~\ref{sec:review_of_np}.
The \MA\ and also \QCMA\ settings elicit a new challenge: on ``yes'' instances, there may be an exponentially larger number of (classical) witnesses in the 
gap-interval (e.g., $(1/3,2/3)$) than in the ``yes'' interval $(2/3,1)$. Thus, randomly choosing the witnesses --- in the spirit of the first attempt in the Valiant-Vazirani construction --- would, with overwhelmingly large probability, fail to choose a unique witness from the ``yes'' interval, and no witnesses from the gap interval.
The main idea behind eliminating this obstacle is to divide the ``gap'' 
interval into polynomially many smaller intervals and to argue that 
in at least one interval, the number of witnesses inside the interval is not much larger than the number of witnesses in the intervals above it (see Observation~\ref{ob:existence_of_lightweight_range} for detail). Therefore, by guessing the approximate sizes of the ``gap'' interval and of all the intervals above it, we have a constant probability that exactly one element from these intervals will be filtered and that this element will not be from the ``gap'' interval.

We defer the discussion about some impossibility result related to $\UQMA$ to Section~\ref{sec:intro_impossibility_uqma}. Here we use the definition of $\UQMA$ together with Theorem~\ref{thm:uqcma_problem_which_is_QCMA_hard} to derive interesting implications. 


\subsection{Implications for Hamiltonian Complexity}
\label{sec:implications_to_hamiltonian_complexity}
We say that a Hamiltonian acting on $n$ $d$-dimensional particles, is $k$-local if it can be written as a sum of $\poly(n)$ terms that act non-trivially on at most $k$ sites.

\begin{definition}
k-$\lh$: We are given a $k$-local Hamiltonian on $n$ qubits $H = \sum_{j=1}^r H_j$ with $r = \poly(n)$. Each $H_j$ is a Hermitian operator with a bounded operator norm $|| H_j || \leq \poly(n)$ acting non-trivially on at most $k$ qudits. We are also given two constants, $a$ and $b$, with $b - a \geq 1/\poly(n)$. In ``yes'' instances, the smallest eigenvalue of $H$ is at most $a$. In ``no'' instances, it is larger than $b$. We should decide whether it is a Yes case or a No case. 
\end{definition} 

In a seminal work, Kitaev showed that the 5-$\lh$ problem is complete for $\QMA$ \cite{Kit99,KSV02}. Improvements in parameters (dimensionality and locality) were given in 
\cite{KR03,KKR06,OT10}, leading to the $\QMA$-completeness of the \onedlh\ \cite{AGIK07,HNN13}, which is the variant of the original problem to one-dimensional nearest-neighbor Hamiltonians (where the local dimension of every qudit, according to \cite{HNN13} can be as low as $d=8$). The 
importance of these results stems not only from the fact 
that $\lh$ is probably the most representative $\QMA$-complete problem, 
but also from the key role played by local Hamiltonians and their ground-state energy in physics. 

An important parameter when dealing with the complexity of 
ground states and local Hamiltonians  
is the \textit{spectral gap} of local Hamiltonians, which is
given by the difference between the ground and the first excited 
energy levels, $\Delta := \lambda_{1}(H) - \lambda_0(H)$. From now on in the discussion, we assume that the operator norm of each term in the Hamiltonian is bounded by some constant and that there cannot be two terms acting on the same set of qubits. 
When the spectral gap is constant, the Hamiltonian is said to be 
gapped. When it is inverse polynomial, we say the Hamiltonian is 
poly-gapped. 

What are the implications of a gap for the $\lh$ problem? 
A groundbreaking result by Hastings shows that ground states of
1-D gapped Hamiltonians have an efficient classical approximation by a Matrix-Product-State (MPS) of polynomial bond dimension \cite{Has07}\footnote{A state $\ket{\psi} \in (\mathbb{C}^d)^{\otimes n}$ has an MPS representation with bond dimension $D$ if it can be written as
\begin{equation}
\ket{\psi} = \sum_{i_1,...,i_n=1}^d \trace(A_{i_1}^{[1]}...A_{i_n}^{[n]}) \ket{i_1,...,i_n},
\end{equation}
with $A_i^{[k]}$ $D \times D$ matrices. Note that only $ndD^2$ complex numbers are needed to specify the state.}. Since the expectation values of local observables of an MPS can be calculated in time polynomial in the number of sites and in its bond 
dimension (see e.g., \cite{PVWC07}), Hastings' 
result implies that $\onedcgulh$ (the restriction of the original problem to 1-D gapped Hamiltonians) belongs to $\NP$.\footnote{Hastings' result was extended in various directions. In particular, a classical algorithm that efficiently finds the MPS approximation of the ground state was shown to exist, see~\cite{LVV15,ALVV17}.}

The question of whether such efficient descriptions might also exist for the ground state of 1-D poly gapped Hamiltonians has also been asked. Under some reasonable complexity-theoretic assumption, we answer this question in the negative. The reasoning for our negative answer is as follows.

We define the $\ulh$ problem as similar to the $\lh$  problem, 
where the conditions  for a ``no'' instance are the same, but 
for a ``yes'' instance we demand that there exists 
a state $\ket{\psi}$ with energy below the lower threshold 
and all other eigenvalues above the upper threshold. 
We similarly define the $\onedulh$. 

The $\onedlh$ is $\QMAC$~\cite{AGIK07,HNN13}. We show that a similar result holds for the ``unique'' variant of this problem. 
\begin{theorem}
The\label{thm:onedulh_is_uqmac}
 \onedulh\  problem\ is $\UQMAC$.
\end{theorem}
The main observation here is that the construction of Aharonov et al. preserves the uniqueness. The precise definition of the problem as well as the proof are given in Section \ref{sec:marriott-watrous}, p.~\pageref{pf:onedulh_is_uqmac}.
Combining Theorems~\ref{thm:onedulh_is_uqmac} and \ref{thm:uqcma_problem_which_is_QCMA_hard}, we have:
\begin{corollary}
The\label{cor:onedulh_is_qcma_hard} 
\onedulh\  problem\ is $\QCMA$-hard under randomized reductions.
\end{corollary}

From Corollary \ref{cor:onedulh_is_qcma_hard}, we can deduce the following no-go result for the ground state of poly-gapped Hamiltonians. Consider any set of states that 
are (i) described by $\poly(n)$ parameters and (ii) from which one can 
efficiently compute expectation values of local 
observables. Matrix product states consititute an example of such a set, and several 
others have recently been proposed 
\cite{APDV+06,Vid07,HKHD+09}. We can show: 

\begin{corollary}
\label{corapprox} 
If all ground states of 1-D poly gapped local Hamiltonians can be approximated to inverse polynomial accuracy by states satisfying properties (i) and (ii) above, then  
$\QCMA \subseteq \RP^\NP$.\footnote{Recall that $\RP$ is defined as $\BPP$, except every $x$ not in the language needs to be rejected with certainty. In other words, languages in $\RP$ have perfect soundness.}
\end{corollary}
Since it seems unlikely that $\QCMA \subseteq \RP^\NP$, we view this as a no-go corollary.

\begin{proof-sketch}
We will show that under the assumptions of the corollary, $\onedulh \in \NP$. Combining that with Corollary~\ref{cor:onedulh_is_qcma_hard}, by Observation~\ref{ob:proerties_of_randomized_reductions}.\ref{it:A_is_C_hard_and_A_in_D} we get the desired result, i.e., $\QCMA \subseteq \RP^\NP$.

Consider the following $\NP$ verification for the $\onedulh$ problem. The prover sends a \emph{classical} witness, which approximates the ground-state with properties (i) and (ii) above, to the verifier. The length of the witness is polynomial in $n$ by property (i). The verifier uses property (ii) to efficiently calculate the expectation value of the local Hamiltonian $H$ that was received as the input. The verifier accepts if the energy is at most $\frac{a+b}{2}$, and rejects otherwise. Completeness and soundness hold by construction, and therefore, $\onedulh\in \NP$.
\end{proof-sketch}


To further analyze the complexity of the local Hamiltonian problem for poly-gapped Hamiltonians, we introduce a variant of the $\UQMA$ class, which we call the poly-gapped $\QMA$ (\PGQMA), as follows: in both ``yes'' and ``no'' instances, we require that there be a gap (given by a pre-determined quantity larger than an inverse polynomial in the input size) from the witness, which accepts with the largest probability, to all the others. We show that the problem \onedpglh, in which the Hamiltonians are promised to be poly-gapped, is $\PGQMAC$ (see Theorem~\ref{thm:onedpglh_is_PGQMAC}). We also present a simple randomized reduction from any $\UQMA$ problem to a $\PGQMA$ --- see Lemma~\ref{le:pgqma_equals_uqma_randomized} --- which is used to show that:

\begin{theorem} 
\label{thm:onedplh_is_qcma_hard} 
The \onedpglh\  problem\ is $\QCMA$-hard under randomized reductions.
\end{theorem}
The proof is given on p.~\pageref{pf:onedplh_is_qcma_hard}. We thus see that, unless $\BQP = \QCMA$ \footnote{$\BQP$ is the class of problems that can be efficiently solved with high probability, by a quantum computer}, the determination of the ground energy of poly-gapped 1-D local Hamiltonians is an intractable problem for quantum computation. Note that this conclusion cannot be drawn from the previous lower bounds on the complexity of the problem \cite{AGIK07, SCV08}. Indeed, the results of \cite{AGIK07} concerning adiabatic quantum computation with a 1-D poly-gapped Hamiltonian indirectly imply that the $\onedpglh$ problem is $\BQP$-hard\footnote{The construction of \cite{AGIK07} for adiabatic quantum computation with one-dimensional Hamiltonians provides a way to encode the outcome of any polynomial quantum computation into the expectation value of a measurement, in the computational basis, of the first site of the ground state of a 1-D poly-gapped local Hamiltonian with a zero ground state energy. By adding a small perturbation to the Hamiltonian, penalizing the first site when it is not in the zero state and with a strength that is much smaller than the spectral gap but that is still inverse polynomial in the number of sites, we can readily conclude that this construction shows that the $\onedpglh$ problem is $\BQP$-hard}, while in \cite{SCV08}, the problem was shown to be hard for the class $\UNP \cap \coUP$ (the intersection of unique $\NP$ with its complement), whose relation with $\BQP$ is unknown.  

\subsection{Impossibility Results for \UQMA\ }
\label{sec:intro_impossibility_uqma}
Finally, we examine the \UQMA\ case.
We show that, when attempting to apply the brute force analogue of 
the previous proofs in the case of $\UQMA$, we already fail in 
the first (inefficient) component. A new idea seems to be required if an extension of the Valiant-Vazirani approach is possible at all for $\QMA$. 

This challenge is demonstrated by a simple family of $\QMA$ ``yes'' instances for which the first component fails to work:
\begin{example}
\label{ex:qma_2_accepting_states}
Let $C$ be a quantum circuit on $l$ qubits with the property that there exists a subspace $V$ of dimension 2, s.t. $\forall \ket{\psi}\in V,\ \Pr(C\ accepts\ \ket{\psi})=1$, and $\forall \ket{\psi}\in V^\bot,\ \Pr(C\ accepts\ \ket{\psi})=0$.
\end{example}

In the classical case, the analogous example of two solutions 
is easy to deal with by choosing a ``filter'' (hash-function) that screens about half of 
the witnesses. A suitable natural quantum analogue is to use a random projection that will reject half of the space. In proposition \ref{failQMA}, we prove that such a transformation does not create an inverse polynomial gap between the two states in the subspace 
$V$: with probability exponentially close to 1, regardless of the dimension of the random projection, all states in $V$ are accepted with probabilities exponentially close to each other. 



The reason for this is that the projection of every $N$-dimensional vector on a $d$-dimensional random subspace is concentrated around $\frac{d}{N}$, with a standard deviation of order $\frac{\sqrt{d}}{N}$ for a sufficiently large $N$. Therefore, regardless of how we choose $d$, we always get that the gap is less than $\frac{1}{\sqrt{N}}$ (which is exponentially small in the number of qubits). Hence, the behavior of random sets --- the filters in the classical setting --- is very different from the behavior of random subspaces, the natural quantum analogue.



One might hope that a more refined measurement would help. In fact, \cite{Sen06} has shown that the two distributions that result from applying a random von Neumann measurement on two arbitrary orthogonal states have a constant total variation distance with all but exponentially small probability. This sounds promising; moreover, a similar effect can be achieved efficiently
by quantum $t$-designs as shown by \cite{AE07}.
Unfortunately, a constant total variation distance between two distributions does not
imply an efficient method to distinguish between them; this problem
is tightly related to complete problems for the complexity class $\SZK$, 
which are not known to have a quantum polynomial time algorithm.
Thus, the problem of whether there exists a randomized reduction from $\QMA$ to $\UQMA$ remains open.

\subsection{Subsequent Works}
Since the first version of this work appeared in 2008~\cite{ABBS08}, several other papers have extended the study of notions related to those that constituted the focus of this work. 

In Section~\ref{sec:intro_impossibility_uqma}, we presented Example~\ref{ex:qma_2_accepting_states}, for which the direct approach to show a reduction from $\QMA$ to $\UQMA$ fails. We argued that an alternative idea is needed. Indeed, in~\cite{JKK+12}, Jain et al. presented a different approach that successfully, among other things, tackles the example given there. Let $\FewQMA$ be the analogue of $\UQMA$ with polynomially many witnesses. They show that $\FewQMA \subseteq \P^{\UQMA}$ with the following alternative technique. Another way to resolve the classical analog in Example~\ref{ex:qma_2_accepting_states} (i.e., an $\NP$ instance with exactly two accepting witnesses) is to ask for $2$ witnesses in lexicographic order that both pass the original verification. Indeed, if the valid witnesses are $w_1$ and $w_2$ (where we assume $w_1<w_2$), the unique witness that will be accepted is $(w_1,w_2)$. The approach can be generalized to the setting in which there are (at most) polynomially many witnesses. But what could take the analog of the lexicographic order in the quantum setting? It turns out that the properties of the anti-symmetric subspace bear some resemblance to those of the lexicographic order. For example, suppose the subspace $V$ in Example~\ref{ex:qma_2_accepting_states} is spanned by the orthogonal basis $\ket{\psi_1}$ and $\ket{\psi_2}$. In this case, the unique anti-symmetric state with respect to two registers is $\frac{1}{\sqrt 2}(\ket{\psi_1}\otimes \ket{\psi_2}- \ket{\psi_2}\otimes \ket{\psi_1})$. Therefore, the verification that takes two registers and tests that they are both in $V$, and that these two registers are anti-symmetric will have a unique eigenvector that is accepted with probability $1$, while all states orthogonal to it are rejected with certainty. Similar to the classical setting, Jain et al. generalizes  this approach (only) to polynomially many witnesses and prove that $\FewQMA\subseteq \P^\UQMA$.

It is known that the class $\MA$ can have perfect completeness, i.e., $\MA=\MA_1$~\cite{ZF87}. Jordan et al.~\cite{JKNN12} proved an analogous statement for $\QCMA$, i.e., $\QCMA=\QCMA_1$. It remains to be seen whether Theorem~\ref{thm:uqcma_problem_which_is_QCMA_hard} could be strengthened to show there exists a $\UQCMA_1$ promise problem that is $\QCMA\textsf{-Hard}$ (or $\QCMA_1\textsf{-Hard}$, since these classes are equal).

Several other works studied the role of the spectral gap of local Hamiltonians. 
Ambainis~\cite{Amb14} defined the $\spectralgap$ problem: the input is a local Hamiltonian and a parameter $\epsilon$, where $\epsilon$ is inverse polynomial in the number of qubits; the problem is to determine whether the spectral gap of the Hamiltonian is at most $\epsilon$ or above $2\epsilon$. 
Ambainis proved that this problem is in $\P^{\QMA[log]}$,  that is, a polynomial TM with logarithmic number of oracle queries to $\QMA$. Gharibian and Yirka~\cite{GY19} proved that it is $\P^{\UQMA[log]}\textsf{-hard}$ under a Cook reduction. Gharibian and Yirka improved Ambainis result and proved hardness\footnote{Gharibian and Yirka identified a problem in Ambainis' proof. Whereas their hardness proof is under a Cook reduction, Ambainis' proof used a Karp reducion.} for $4$-local Hamiltonians.  
González-Guillén and Cubitt~\cite{GGC18} proved that it is impossible to show the $\QMA$-hardness of \emph{constant} gapped Hamiltonians via certain generalizations of Kitaev's circuit-to-Hamiltonian construction (see~\cite[Section 14.4.1]{KSV02}). It was shown that the problem of deciding whether a translationally invariant Hamiltonian has a constant gap or is gap-less, when taking the size of the system $n$ to infinity, is undecidable on a 2-D ~\cite{CPGW15} or a 1-D~\cite{BCLPG18} system.
Fefferman and Lin proved that $\PreciseQMA=\PSPACE$, where the completeness and soundness gap in $\PreciseQMA$ can be exponentially small~\cite{FL16}. Recently, Deshpande, Gorshkov and Fefferman~\cite{DGF22} defined the class $\PrecisePGQMA$, which restricts $\PreciseQMA$ to instances that have an inverse-polynomial gap, and proved that $\PrecisePGQMA=\PP$. The spectral gap also plays a role in quantum algorithms. For example, in Ref.~\cite{GS17}, an algorithm that constructs the ground-state of a special class of Hamiltonians. The running time of the algorithm scales inverse polynomially with the \emph{uniform} spectral gap of the Hamiltonian, meaning the minimal spectral gap of every subsystem -- see~\cite{GS17} for more detail.

\subsection{Organization}
The structure of the remainder of the paper is as follows: in Section~\ref{sec:definitions}, we present the definitions. Section \ref{sec:review_of_np} comprises a review of the proof of the Valiant-Vazirani Theorem, while Sections \ref{sec:ma} and \ref{sec:vv_to_qcma} contain the extensions of the theorem to the classes $\MA$ and $\QCMA$, respectively. In Section \ref{sec:robustness}, we discuss some alternative definitions of the class \UQMA and complete problems for this class. We also show that the two classes are equivalent under randomized reductions.
Finally, in Section \ref{sec:qma_impossibility_results}, we prove the impossibility results for the extension of our results
to \QMA\ using similar ideas. 
\section{Definitions}
\label{sec:definitions}
This work lies at the intersection of three fields with which we assume the reader has some familiarity: Computational complexity (see~\cite{AB09}), Quantum Computing (see~\cite{NC16}), and Hamiltonian Complexity (see~\cite{GHLS15}). We begin by defining a few standard complexity classes that we will consider throughout the paper. Then we turn to the definition of unique versions of $\MA$, $\QCMA$, and $\QMA$ that, to the best of our knowledge, have not been formalized before. 

\subsection{Preliminary Definitions}
\label{sec:preliminary_definitions}


\begin{definition}[Nondeterministic Polynomial Time(\NP)\cite{Coo71}]
A language $L\in \NP$ if there exists a Turing Machine (TM) $M$ that runs in polynomial time in its first argument s.t. for every $x \in \{0,1\}^*$:
\begin{enumerate}
\item $x\in L \Rightarrow \exists y\ s.t.\ M(x,y)\ accepts$.
\item $x \notin L \Rightarrow \forall y$ $M(x,y)\ rejects$.
\end{enumerate}
\end{definition}

\begin{definition}[Randomized Polynomial Time (\RP)\cite{Gil77}]
A language $L\in \RP$ if there exists a randomized Turing Machine (TM) $M$ that runs in polynomial time s.t. for every $x \in \{0,1\}^*$:
\begin{enumerate}
\item $x\in L \Rightarrow \Pr(M(x)\text{ accepts})\geq \frac{2}{3}$.\footnote{The probability is over the coin tosses of the randomized TM.}
\item $x \notin L \Rightarrow \Pr(M(x) \text{ accepts}) = 0$.
\end{enumerate}
\end{definition}

\begin{definition}[{Randomized reduction (adapted from \cite{VV85})}]
A promise problem $\textsc{A}$ is reducible to a promise problem $\textsc{B}$ by a randomized reduction, if there exists a randomized polynomial Turing Machine (TM) $M$ and a polynomial $p$ s.t.:
\begin{itemize}
\item Completeness: $x\in \textsc{A}_{yes} \Rightarrow \Pr_y(M(x,y) \in \textsc{B}_{yes})\geq 1/p(|x|)$
\item Perfect soundness\footnote{Some texts, e.g.,~\cite[Section 7.6]{AB09}, require that completeness and soundness hold with probability $\frac{2}{3}$. We use the more stringent definition, which allows us to use the facts in Observation~\ref{ob:proerties_of_randomized_reductions}. None of these properties (including transitivity!) hold if the reduction does not have perfect soundness.}: $x\in \textsc{A}_{no} \Rightarrow \forall y \  M(x,y) \in \textsc{B}_{no}$,
\end{itemize}
where $y$ are the random bits of the TM $M$. We denote this by $\textsc A \leq_r \textsc B$. We say that a promise problem $\textsc B$ is $\CH$ under randomized reductions for some complexity class $\C$ if for every promise problem $A\in \C$, $A\leq_r \textsc B$.  
\label{def:randomized_reduction}
\end{definition}

The motivation behind randomized reductions stems from, among other reasons, the following immediate properties:
\begin{observation}
\begin{enumerate}
    \item \label{it:random_reduciton_transitive} If $\textsc A_1 \leq_r \textsc A_2$ and $\textsc A_2 \leq_r \textsc A_3$, then $\textsc A_1\leq_r \textsc A_3$. 
    \item \label{it:A_is_CH_and_A_reducible_B_implied_B_C_hard} If $\textsc A$ is $\CH$ (or even $\CH$ under randomized reductions) and $\textsc A \leq_r \textsc B$, then $\textsc B$ is $\CH$ under randomized reductions. 
    \item \label{it:A_reducible_B_and_B_in_RP} If $\textsc{A} \leq_r \textsc B$ and $\textsc{B} \in \RP$, then $\textsc A\in \RP$. 
    \item \label{it:A_is_C_hard_and_A_in_D} If $\textsc A$ is $\CH$ under randomized reductions, and $\textsc A \in \textsf{D}$ then $\C \subseteq \RP^{\D}$.
    \item \label{it:A_is_C_hard_and_A_in_D_and_B_is_D_hard_then_B_is_C_hard_random} If $\textsc A$ is $\CH$ under randomized reductions, and $\textsc A \in \textsf{D}$ and $\textsc B$ is $\DH$ (or, even $\DH$ under randomized reductions\footnote{Here we use the transitivity of randomized reductions --- see Item~\ref{it:random_reduciton_transitive}.}) then $\textsc B$ is also $\CH$ under randomized reductions.
    \item \label{it:A_is_C_hard_and_A_in_RP} If $\textsc A$ is $\CH$ under randomized reductions, and $\textsc A \in \RP$ then $\C \subseteq \RP$.
\end{enumerate}
\label{ob:proerties_of_randomized_reductions}
\end{observation}

\begin{definition}[{Unique Nondeterministic Polynomial Time (\UNP\footnotemark)\cite{Val76}}]
\footnotetext{Due to Valiant, the original abbreviation stands for ``Unambiguous Polynomial''.}A promise problem $L=(L_{yes},L_{no}) \in \UNP$ if there exists a Turing Machine (TM) $M$ that is polynomial in its first argument s.t. for every $x \in \{0,1\}^*$:
\begin{enumerate}
\item $x\in L_{yes} \Rightarrow \exists y\ s.t.\ M(x,y)\ accepts$ and $\forall y'\neq y\ M(x,y')\ rejects$.
\item $x \in L_{no} \Rightarrow \forall y$ $M(x,y)\ rejects$.
\end{enumerate}
\label{def:unp}
\end{definition}
\begin{definition}[Merlin-Arthur (\MA)\cite{Bab85}]
A promise problem $L=(L_{yes},L_{no}) \in \MA$ if there exists a probabilistic polynomial TM $M$ that is polynomial in its first argument, and its random bits are denoted by the string $r$, s.t. for every $x \in \{0,1\}^*$:
\begin{enumerate}
\item $x\in L_{yes} \Rightarrow \exists y\ s.t.\ \Pr_r(M(x,y,r)\ accepts)\geq 2/3$.
\item $x \in L_{no} \Rightarrow \forall y$ $\Pr_r(M(x,y,r)\ accepts)\leq 1/3$.
\end{enumerate}
\end{definition}

\begin{definition}[Quantum Classical Merlin-Arthur (\QCMA)\cite{AN02}]
A promise problem $L=(L_{yes},L_{no}) \in \QCMA$ if there exists a uniformly generated\footnote{Uniformity means that the circuit $U_x$ could be deterministically computed from $x$ in polynomial time.} quantum circuit $U_x$, having $l(x)$ qubits as input and requiring $m(x)$ ancilla qubits initialized to $\ket{0^m}$, such that for every $x \in \{0,1\}^*$
\begin{enumerate}
 \item $x\in L_{yes} \Rightarrow \exists y\ s.t. \ \lVert \Pi_1 U_x(\ket{y}\otimes\ket{0^m}) \rVert^2 \geq 2/3$.
\item $x\in L_{no} \Rightarrow \forall y \ \lVert \Pi_1 U_x(\ket{y}\otimes \ket{0^m})\rVert^2 \leq 1/3$.
\end{enumerate}
$\Pi_1$ is the projection onto $\ket{1}$ in the first qubit, i.e., $\Pi_1 := \ketbra{1}{1}\otimes I_{l+m-1}$. For brevity, we write $l=l(x)$ and $m=m(x)$ when $x$ can be understood from the context. 
\end{definition}

\begin{definition}[Quantum Merlin-Arthur (\QMA)\cite{Kit99,KSV02}] \label{QMAdef}
A promise problem $L=(L_{yes},L_{no}) \in \QMA$ if there exists a uniformly generated quantum circuit $U_x$, having $l(x)$ qubits as input and requiring $m(x)$ ancilla qubits initialized to $\ket{0^m}$, such that for every $x \in \{0,1\}^*$
\begin{enumerate}
 \item $x\in L_{yes} \Rightarrow \exists \ket{\psi}\ s.t. \ \lVert \Pi_1 U_x(\ket{\psi}\otimes\ket{0^m}) \rVert^2 \geq 2/3$.
\item $x\in L_{no} \Rightarrow \forall \ket{\psi}$  $\lVert \Pi_1 U_x(\ket{\psi}\otimes\ket{0^m}) \rVert^2 \leq 1/3$.
\end{enumerate}
$\Pi_1$ is the projection onto $\ket{1}$ in the first qubit.
\end{definition}

\subsection{New Definitions}
We now provide the analogous, unique versions for the classes $\MA$, $\QCMA$ and $\QMA$. 

\begin{definition}[Unique Merlin-Arthur (\UMA)]
A promise problem $L=(L_{yes},L_{no}) \in \UMA$ if there exists a probabilistic TM $M$ that is polynomial in its first argument s.t. for every $x \in \{0,1\}^*$:
\begin{enumerate}
\item $x\in L_{yes} \Rightarrow \exists y \text{ s.t. } \Pr_r(M(x,y,r)\ accepts)\geq 2/3$ and \hbox{$\forall y'\neq y$, $\Pr_r(M(x,y',r)\ accepts)\leq 1/3$}.
\item $x \in L_{no} \Rightarrow \forall y \quad \Pr_r(M(x,y,r)\ accepts)\leq 1/3$.
\end{enumerate}
\end{definition}

\begin{definition}[Unique Quantum Classical Merlin-Arthur (\UQCMA)]
A promise problem $L=(L_{yes},L_{no}) \in \UQCMA$ if there exists a uniformly generated polynomial quantum circuit $U_x$ that can be computed in $poly(|x|)$ time, having $l(x)$ qubits as input and requiring $m(x)$ ancilla qubits initialized to $\ket{0^m}$, such that for every $x \in \{0,1\}^*$,
\begin{enumerate}
 \item $x\in L_{yes} \Rightarrow \exists y\ s.t. \ \lVert \Pi_1 U_x(\ket{y}\otimes\ket{0^m}) \rVert^2 \geq 2/3$ and $\forall y'\neq y$, $\lVert \Pi_1 U_x(\ket{y'}\otimes\ket{0^m}) \rVert^2 \leq 1/3$
\item $x\in L_{no} \Rightarrow \forall y \ \lVert \Pi_1U_x(\ket{y}\otimes\ket{0^m}) \rVert^2 \leq 1/3$.
\end{enumerate}
$\Pi_1$ is the projection onto $\ket{1}$ in the first qubit.
\end{definition}

\begin{definition}[Unique Quantum Merlin-Arthur (\UQMA)]
\label{def:uqma}
A promise problem $L=(L_{yes},L_{no}) \in \UQMA$ if there exists a uniformly generated polynomial quantum circuit $U_x$ that can be computed in $poly(|x|)$ time, having $l(x)$ qubits as input and requiring $m(x)$ ancilla qubits initialized to $\ket{0^m}$, s.t.
\begin{enumerate}
 \item $x \in L_{yes} \Rightarrow \exists \ket{\psi} \lVert \Pi_1 U_x(\ket{\psi}\otimes\ket{0^m}) \rVert^2 \geq 2/3 \hspace{0.1 cm} \text{and} \hspace{0.1 cm} \forall \ket{\varphi}\bot \ket{\psi}, \hspace{0.05 cm} \lVert \Pi_1 U_x(\ket{\varphi}\otimes\ket{0^m}) \rVert^2 \leq 1/3$ 
\item $x\in L_{no} \Rightarrow \forall \ket{\psi}$ $\lVert \Pi_1 U_x(\ket{\psi}\otimes\ket{0^m}) \rVert^2 \leq 1/3$.
\end{enumerate}
\end{definition}
Later we show an alternative way to define this class, see Definition~\ref{def:uqma_watrous} and Lemma~\ref{lem:uqma_definitions_are_equivalent}.

\section{The Valiant-Vazirani Theorem Revisited}

\label{sec:review_of_np} 
In the introduction, the Valiant-Vazirani theorem was mentioned as something that should be interpreted as ``it is not much easier to solve $\UNP$ than it is to solve $\NP$''. We now discuss this interpretation in greater detail.
By combining the Valiant-Vazirani theorem (see Theorem~\ref{thm:valiant-vazirani}) with Observation~\ref{ob:proerties_of_randomized_reductions}.\ref{it:A_is_C_hard_and_A_in_D} (where we use $\D=\UNP$ and $\C=\NP$), we obtain the following conclusion:
\begin{corollary}
$\NP\subseteq \RP^\UNP$.
\label{cor:np_subseteq_rp_up}
\end{corollary}
In other words, an efficient randomized algorithm for a $\UNPC$ problem implies an efficient randomized algorithm for all of $\NP$. Of course, it is conjectured\footnote{Recall that it is conjectured that $\RP=\BPP=\P$~\cite{IW97} (see also~\cite[Chapter 20]{AB09}). Therefore, $\UNP\subseteq \RP$, together with the previous conjecture violates the (much more famous and established) $\P\neq \NP$ conjecture.} that no such efficient randomized algorithm exists for $\NP$, and therefore, such an efficient algorithm could not exist for a $\UNPC$ problem. Our results could also be used to show impossibility results based on natural complexity conjectures --- see, for example, Corollary~\ref{corapprox}.

We are now ready to review the proof of the Valiant-Vazirani theorem (see Theorem~\ref{thm:valiant-vazirani}). We divide the proof into three components, so that we can better understand which components of the original construction fail in the probabilistic and quantum settings. 

The original proof of the theorem works with the well known $\NP$-complete problem $\sat$. We will not use it, however, because $\sat$ lacks a simple variant that is complete for the classes $\MA$ or $\QCMA$. Instead, we will use the following problem:
\begin{definition}[\textsc{Trivial NP Problem} (\tnpp)]
The words in $\tnpp$ are tuples, $\langle V,x,l,t\rangle$, where V is a description of a deterministic Turing machine, x is a string, and $l,t\in \mathbb{N}$, given in unary.

$\langle V,x,l,t\rangle\in \tnpp$ iff there exists a $y\in\{0,1\}^l$ s.t. $V(x,y)$ accepts in at most $t$ steps. 
\label{def:tnpp}
\end{definition}
It can easily be seen that $\tnpp$ is $\NPC$. Similarly, the following promise problem is a ``unique'' version of \tnpp that is $\UNPC$.

\begin{definition}[\textsc{Unique-NP Promise Problem (\unppp)}]
The promise problem  $\unppp=(\unppp_{yes},\unppp_{no})$. The words in $\unppp$ are tuples, $\langle V,x,l,t\rangle$, where V is a description of a deterministic Turing machine, x is a string of length n, and $l,t\in \mathbb{N}$, given in unary.
 
$\langle V,x,l,t\rangle\in \unppp_{yes}$ if there exists exactly one string $y\in \{0,1\}^l$ s.t. $V(x,y)$ accepts in at most $t$ steps.
$\langle V,x,l,t\rangle\in \unppp_{no}$ if for all strings $y\in \{0,1\}^l$, $V(x,y)$ does not accept in $t$ steps.
\end{definition}

\subsection{Proof Sketch}
Our goal is to prove Theorem~\ref{thm:valiant-vazirani} by showing that $\unppp$ is $\NPH$ under randomized reductions. This completes the proof since, as mentioned already, $\unppp\in \UNP$. Since $\tnpp$ (recall Definition~\ref{def:tnpp}) is $\NPH$, by Observation~\ref{ob:proerties_of_randomized_reductions}.\ref{it:A_is_CH_and_A_reducible_B_implied_B_C_hard}, it is enough to prove that $\tnpp \leq_r \unppp$. 

We present the proof in a series of attempts, each of which improves on its predecessor by introducing a new component. The third attempt is that which completes the proof.

\subsubsection{Component 1: The right random ``filter'' for the right size} 

For a $\tnpp$ instance $I=\langle V,x,l,t\rangle$, let $W$ be its set of accepting witnesses:  
\[ W \equiv \lbrace y \in \{0,1\}^l :   V(x,y) \text{ accepts in at most  $t$\ steps} \rbrace,\]
and let $w\equiv |W|$. Notice that $I\in \tnpp \iff w\neq 0$. 


\begin{definition}[R-restriction]
\label{def:r-restriction}
Let $R$ be a set of strings of length $l$, with the property that there is an algorithm that answers whether $y\in R$ in exactly $T$ time steps. Given a Turing machine $V$, we call the following Turing machines the $R$-restriction of V and denote it by $V_R$:
\begin{enumerate}
\item If $y \notin R$, Reject. Otherwise, Continue.
\item Run $V$ on $(x,y)$.
\end{enumerate} 
\end{definition}

We view the $R$-restriction as a filter added to the original problem, because the new machine, $V_R$, accepts only accepting witnesses of the original machine $V$, which belong to the set $R$. 

Let us denote by $I'$ the instance $\langle V_R,x,l,t+T\rangle$. Component 1 takes the filter $R$ to be a random set, where each string in $\{0,1\}^l$ is chosen independently with probability $\frac{1}{w}$. Notice that the Turing machine $V_R$ may not have a short description, because to decide whether $y\in R$, all the elements of $R$ should somehow be ``hard-wired'' to the machine. Of course, using Kolmogorov complexity  arguments~\cite{CT06}, one can easily show that for large values of $|R|$, the length of the description of the Turing machine that decides whether $y\in R$ cannot be polynomial. Therefore, the mapping between $I$ to $I'$ is not efficient. Similarly, $T$, which was defined as the time it takes for the TM to decide membership in $R$, might not be polynomial. These drawbacks will be circumvented in component 3. 

We claim that $I'$ will be in $\unppp_{yes}$ with probability $\Omega(1)$. Let $W'= \lbrace y\in \{0,1\}^l : \  V_R(x,y)\ \text{accepts in}\ t+T\ \text{steps} \rbrace$. Defining $W = \{y_1,...,y_w\}$, 

\begin{align}
\Pr\left (I'\in \unppp_{yes}\right ) &= \Pr \left (|W'|=1 \right ) \nonumber \\
&= \Pr \left(|W\cap R|=1\right) \nonumber \\
&= \Pr\left( \bigcup_{i=1}^w\left( y_i\in R \cap_{j\neq i} y_j\notin R\right)\right) \nonumber \\
&= \sum_{i=1}^w \Pr\left(  y_i\in R \cap_{j\neq i} y_j\notin R\right) \nonumber \\
&= w \frac{1}{w} \left(1-\frac{1}{w}\right)^{w-1} \nonumber \\ &\geq 1/e.
\label{eq:Iprime_is_yes_with_constant_probability}
\end{align}

The first equality follows from  $I'\in \unppp_{yes} \iff |W'|=1$ and the second from $W'=W\cap R$. The third is a direct consequence of the definition of $y_i$. The fourth stems from the fact that the events in the line above are all disjoint. The fourth is based on the construction of $R$ (where each element was added to the set independently with probability $\frac{1}{w}$).  Therefore,  $\langle V',x,l,t+t'\rangle$ is a ``yes'' with probability of at least $\frac{1}{e}$.

Using this idea, we define $2^l$ (random) instances, $I_1,...,I_{2^l}$, one for every possible value of $w$: $I_j=\langle V_{R_j},x,l,t+T_j\rangle$. Here, $R_j$ is sampled so that each string belongs to $R_j$ with probability $\frac{1}{j}$, and $T_j$ is the running time of the algorithm that decides membership in the set $R_j$ (note that $T_j$ may be exponential). We claim:

\begin{lemma}
\label{le:completeness_and_soundness}
(Completeness) If $I \in \tnpp$, then there exists a $j\in [2^l]$ for which, with probability $\Omega(1)$ over the choice of $R$, $I_j \in \unppp_{yes}$. (Soundness) If $I \notin \tnpp$, then all the $I_j$ are in $\unppp_{no}$.
\end{lemma}

\begin{proof}
The completeness follows from the previous argument: one of the $I_j$'s is $I_w$. By Eq.~\eqref{eq:Iprime_is_yes_with_constant_probability}, $I_w \in \unppp_{yes}$ with probability of at least $1/e$. Soundness: $I\notin \tnpp \Rightarrow W=\emptyset$. As $W_j=W\cap R_j$, $W_j=\emptyset$, and therefore, $I_j \in \unppp_{no}$. 
\end{proof}

Suppose we try to prove that $\tnpp\leq_r \unppp$ by using Lemma~\ref{le:completeness_and_soundness}. 
Consider the following reduction: we sample $j$ from $1$ to $2^n$ uniformly at random and output $\unppp$  $I_j$ as defined above. The completeness asserts that for a ``yes'' instance, we accept with probability $\Omega(1)$ if we guessed $j$ correctly. The soundness asserts that we always reject in a ``no'' instance. We have two sources for the failure for the reduction: (i) the probability to guess $j$ correctly is $\frac{1}{2^l}$, and therefore, completeness holds only with exponentially small probability and (ii) since the sets $R_j$ are chosen uniformly at random, the description of $I_j$ is inefficient and the running time $T_j$ might be exponential (recall that $T$ is given in unary form). Hence, the reduction would take exponential time. In component 2, we resolve the first issue, and in component 3, we resolve the second issue.

\subsubsection{Component 2: An approximated ``filter'' also works} 

The second component addresses the fact that we do not know the value $w$, and therefore, the reduction described in component 1 has an exponentially small completeness probability. The key idea  is that having an approximating $w$ by some constant, multiplicative factor only changes the probability of having a unique solution by another constant factor.

More explicitly, we transform our instance $I$ into a polynomial number of random instances: $I_0,I_1,...,I_l$. These instances are formed by choosing random sets $R_k$ again, but now, each element is taken with probability $\frac{1}{2^k}$. 

A similar statement to Lemma \ref{le:completeness_and_soundness} also holds here, despite the fact than now we have only $l$ instances (whereas before we had $2^l$):
\begin{lemma}
\label{le:completeness_and_soundness_component_2}
(Completeness) If $I \in \tnpp$, then there exists a $j\in \{0,\ldots,l\}$ for which, with probability $\Omega(1)$ over the choice of $R$, $I_j \in \unppp_{yes}$. (Soundness) If $I \notin \tnpp$, then all the $I_j$ are in $\unppp_{no}$.
\end{lemma}
\begin{proof}
The soundness analysis follows from exactly the same argument as in component 1 --- see Lemma~\ref{le:completeness_and_soundness}. To analyze the completeness of the protocol, we notice that for some $k\in \{0,\ldots,l\}$, $2^k \leq w < 2^{k+1}$. Hence, for such  $k$, 
\begin{align*}
\Pr\left(I_k \in \unppp_{yes}\right) &= \Pr\left(|W_k|=1\right) \\ 
&=\Pr\left(|W\cap R_k|=1\right) \nonumber \\ 
&= \Pr\left(\bigcup_{i=1}^w\left( y_i\in W \cap_{j\neq i} y_j\notin W\right)\right) \\
&= w \frac{1}{2^k} \left(1-\frac{1}{2^k}\right)^{w-1} \\ 
&\geq \left(1-\frac{1}{2^k}\right)^{2^{k+1}-1}\geq e^{-2}. \qedhere 
\end{align*}
\end{proof}

\subsubsection{Component 3: An approximated pseudorandom filter is just as good}

The third component attends to the inefficiency of the previous reduction: a random and exponential large set $R$ cannot be described by a polynomial description. The solution is to replace the randomness by a suitable notion of pseudorandomness. In this case, the pseudorandom objects of interest are pairwise independent hash functions.
 \begin{definition}[{Pairwise independent hash functions~\cite{CW79}, see also~\cite[Section 8.2.2]{AB09}}]
 \label{def:pairwise_independent_hash_function}
 A family of functions $\mathbb{H}_{n,m}$ where each $h\in \mathbb{H}_{n,m}$, $h:\{0,1\}^n \to \{0,1\}^m$, is called a pairwise independent\footnote{Some texts denote pairwise independent as ``2-universal'' or ``universal$_2$''.} family of hash-functions if:
 
   \[ \forall y_1\neq y_2 \in \{0,1\}^n, \hspace{0.1 cm} \forall a,b\in \{0,1\}^m,\quad  \Pr_{h \sim_{\mathcal{U}} \mathbb{H}_{n,m}}(h(y_1)=a \hspace{0.1 cm} \text{and} \hspace{0.1 cm} h(y_2)= b)=\frac{1}{2^{2m}} \]
 
\end{definition}
Note that this probability is the same as if the map $h$ was chosen uniformly at random from the set of all functions that map $n$ bits to $m$ bits. An interesting property is the existence of families that have concise descriptions and that can be efficiently computed:
\begin{fact}[{~\cite{CW79}, see also~\cite[Section 8.2.2]{AB09}}] Efficient pairwise independent hash functions exist. 
More precisely, there exist a polynomial $T(n,m)$ and a family of pairwise independent hash-functions $\mathbb{H}_{n,m}$ such that
\begin{itemize}
    \item There exists a randomized algorithm to sample a TM that computes $h$ for a uniformly random $h\in\mathbb{H}_{n,m}$. By abuse of notation, we also denote the TM that computes $h$ by $h$. The running time of this algorithm is $T(n,m)$.
    \item For every $h\in \mathbb{H}_{n,m}$, the running time of $h(x)$ is $T(n,m)$.
\end{itemize}
\label{fac:pairwise_indpendent_exist}
\end{fact}
   
In this final attempt, we define $l$ instances $I_1,I_2,\ldots, I_l$. Again, each $I_j$ is based on some $R$-restriction of $V$, but now we use a restriction that is based on an (efficient) pairwise independent hash function: For $j\in \{0,\ldots,l\}$, we sample $h$ from the family $H_{l,j+2}$, and define $R_j=h^{-1}(0)$. We define $I_j=\langle V_{R_j},x,l,t+T(l,j+2)\rangle$. Note that testing membership in $R_j$ is efficient (deciding membership in $R_j$ is done by testing whether $h(y)=0$, which takes $T(l,j+2)$ steps, where $T$ is the polynomial in Fact~\ref{fac:pairwise_indpendent_exist}).

\begin{lemma}
\label{le:completeness_and_soundness_componenet_3}
(Completeness) If $I \in \tnpp$, then there exists a $j\in \{0,\ldots,l\}$ for which, with probability at least $\frac{1}{8}$ over the choice of $R$, $I_j \in \unppp_{yes}$. (Soundness) If $I \notin \tnpp$, then for all $j\in\{0,\ldots,l\}$, $I_j\in \unppp_{no}$. 

Furthermore,  there exists a polynomial time randomized TM that receives as an input $I=\langle V,x,l,t\rangle$ and $j \in \{0,\ldots,l\}$ and that outputs $I_j$.
\end{lemma}
\begin{proof-of-lemma}{\ref{le:completeness_and_soundness_componenet_3}}
Also here, the soundness analysis follows from exactly the same argument as in component 1 --- see Lemma~\ref{le:completeness_and_soundness}.

For completeness, we make use of the following lemma:
\begin{lemma}
\label{le:vv}
Let $W\subseteq \{0,1\}^l$ of size $w$, and let $j \in \NN$ such that  $2^j \leq w < 2^{j+1}$, and let h be a random function sampled from a pairwise independent hash function family $\mathbb{H}_{l,j+2}$. Then,
\begin{equation*}
\Pr \left(|h^{-1}(0)\cap W| = 1 \right) \geq \frac{1}{8}.
\end{equation*}
\end{lemma}

This Lemma is proved in Appendix~\ref{sec:proofs}. Note that $I_j=\langle V_{R_j},l,t+T_{l,j+2}\rangle \in \unppp_{yes}$ is equivalent to $|W_j|=1$. By construction, $W_j=W\cap R_j=W \cap h_j^{-1}(0)$ and using Lemma~\ref{le:vv}, $|h_j^{-1}(0) \cap W|=1$ with probability at least $\frac{1}{8}$ over the choice of $h$. The ``furthermore'' part of the lemma follows immediately from the fact that we use an \emph{efficient} pairwise independent hash-function in the construction of $R_j$.
\end{proof-of-lemma}




We are now ready to prove the Valiant-Vazirani Theorem.

\begin{proofof}{Theorem~\ref{thm:valiant-vazirani}}
The randomized reduction is given in Algorithm~\ref{alg:vv_reduction}.

\begin{algorithm}[H]
\DontPrintSemicolon
  \SetAlgoLined
  \KwIn{ $\langle V,x,l,t\rangle$.}
  Sample $j$ uniformly at random from $\{0,\ldots,l\}$\label{line:sample_k} \;
  Sample $h$ uniformly at random from a family of efficient pairwise independent hash functions $\mathbb{H}_{l,j+2}$, and let $R_j=h^{-1}(0)$\;
  
   \KwOut{$\langle V_{R_j},x,l,t+T(l,j+2)\rangle$}
\caption{The randomized reduction from \tnpp\ to \unppp}
\label{alg:vv_reduction}
\end{algorithm}
The efficiency of the reduction follows from the efficiency of the pairwise-independent hash function that we use --- see the first item in Fact~\ref{fac:pairwise_indpendent_exist}. 
Completeness: with probability at least $\frac{1}{l}$, we sample $j$, which satisfies $2^j\leq w < 2^{j+1}$ in Line~\ref{line:sample_k}; conditioned on guessing $j$ correctly, by Lemma~\ref{le:completeness_and_soundness_componenet_3}, $I_j\in \unppp_{yes}$ with probability at least $\frac{1}{8}$. Overall, if $I\in \tnpp$, then the reduction maps it to a $\unppp_{yes}$ instance with probability at least  $\frac{1}{8l}$. Soundness follows directly from Lemma~\ref{le:completeness_and_soundness_componenet_3}.
\end{proofof}

\section{Valiant-Vazirani Extended to the Class \MA}
\label{sec:ma}

In this section, we prove Theorem \ref{thm:uma_problem_which_is_MA_hard}. Relying on exactly the same argument as in Corollary~\ref{cor:np_subseteq_rp_up}, Theorem~\ref{thm:uma_problem_which_is_MA_hard} implies the following:
\begin{corollary}
$\MA \subseteq \RP^\UMA$. 
\end{corollary}

We first define the promise problems that we will work with throughout this section. 

\begin{definition}[Trivial MA Promise Problem (\tmapp)]
$\tmapp=(\tmapp_{yes},\tmapp_{no})$. The words in $\tmapp$ are tuples, $\langle V,x,p_1,p_2,l,t\rangle$, where V is a description of a randomized Turing machine, $x$ is a string, and $0\leq p_1< p_2\leq 1$, and $l,t\in \mathbb{N}$. The parameters $p_1,p_2,l,t$ are  given in unary\footnote{\label{foot:inverse_poly_gap}The unary representations of $p_1$ and $p_2$ guarantee that $p_2-p_1\geq 1/poly(n)$, where $n$ is the input size.}.
 
$\langle V,x,p_1,p_2,l,t\rangle\in \tmapp_{yes}$ if there exists a string $y$ of length $l$ s.t. $\Pr(V(x,y)\ accepts\ in\ t\ steps) \geq p_2$.

$\langle V,x,p_1,p_2,l,t\rangle\in \tmapp_{no}$ if for all strings $y$ of length $l$, $\Pr(V(x,y)\ accepts\ in\ t\ steps) \leq p_1$.
\label{def:tmapp}
\end{definition}

It can be easily verified that $\tmapp$ is $\MAC$. The containment $\tmapp \in \MA$ uses error reduction for $\MA$ (more precisely, that $\MA_{p_1(n),p_2(n)}\subseteq \MA$, whenever $p_2(n)-p_1(n)\geq 1/poly(n)$ for some polynomial $p$, which is proved in the same way as error reduction for $\BPP$). 

Next, we define the unique variant of $\tmapp$, which is $\UMAC$:
\begin{definition}[Unique MA Promise Problem (\umapp)]
$\umapp=(\umapp_{yes},\umapp_{no})$. The words in $\umapp$ are tuples, $\langle V,x,p_1,p_2,l,t\rangle$, where $V$ is a description of a randomized Turing machine, $x$ is a string, and $0\leq p_1< p_2\leq 1$, and $l,t\in \mathbb{N}$. The parameters $l,t,p_1,p_2$ are  given in unary\footnote{See Footnote~\ref{foot:inverse_poly_gap}.}.
 
$\langle V,x,p_1,p_2,l,t\rangle\in \tmapp_{yes}$ if there exists a string $y$ of length $l$ s.t. $\Pr(V(x,y)\ accepts\ in\ t\ steps) \geq p_2$, and for every $y'\neq y$ of length $l$, $\Pr(V(x,y)\ accepts\ in\ t\ steps) \leq p_1$. 

$\langle V,x,p_1,p_2,l,t\rangle\in \tmapp_{no}$ if for all strings $y$ of length $l$, $\Pr(V(x,y)\ accepts\ in\ t\ steps) \leq p_1$.
\label{def:umapp}
\end{definition}

\begin{figure}[htp]
\centering
\scalebox{0.5}{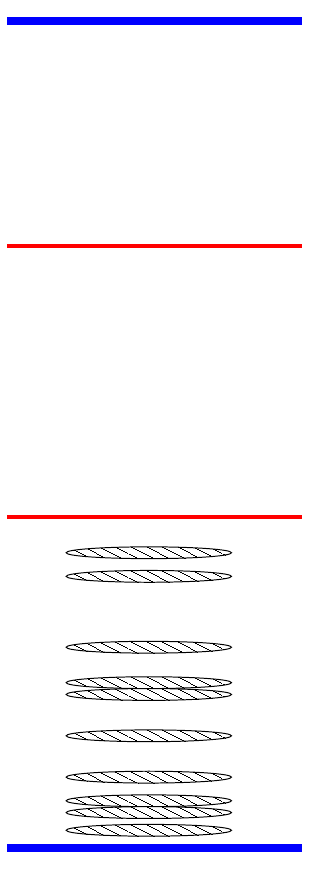}\quad   \scalebox{0.5}{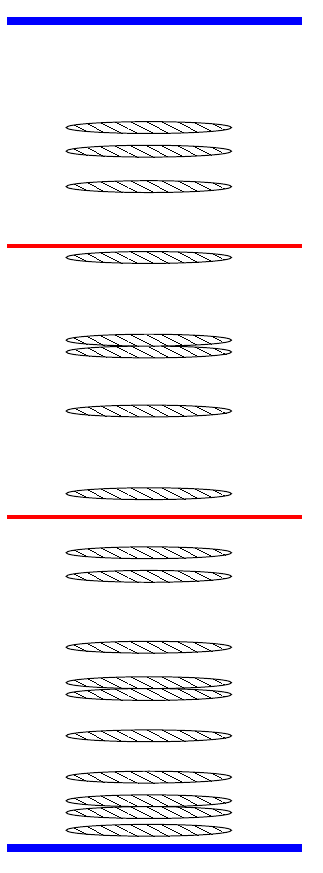}
\caption{Typical ``no'' and ``yes'' $\tmapp$ instances.
The y-axis is probability. The ellipses are all the $2^l$ different witnesses of a specific instance. The red lines outline the boundaries $(p_1,p_2)$ --- the maximal acceptance probability of a $\tmapp$ instance is promised not to be in that interval. The left panel is an example of a ``no'' instance, and the right panel is an example of a ``yes'' instance.}
\label{fig:ma_no_and_yes_instance} 
\end{figure}

\begin{figure}[htp]
\centering
\scalebox{0.5}{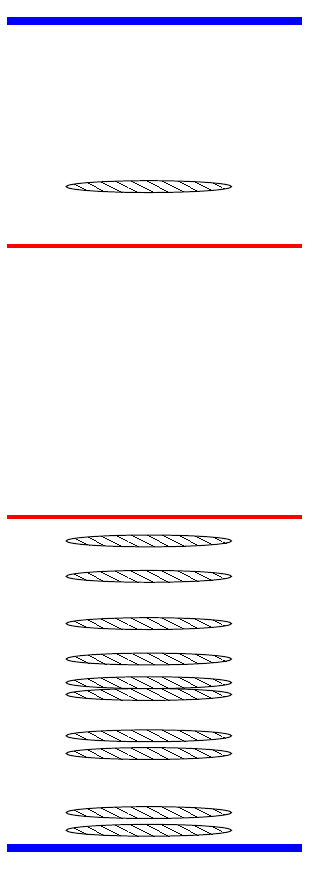}
\caption{A $\umapp_{yes}$  instance. 
There is exactly one witness that is accepted with probability greater than $p_2$, and all others are accepted with probability smaller than $p_1$.}
\label{fig:ma_unique_instance}
\end{figure}

We will prove Theorem~\ref{thm:uma_problem_which_is_MA_hard} by showing that $\tmapp \leq_r \umapp$. Hence, our goal is to create a transformation that takes a $\tmapp_{yes}$ instance (right panel in Fig. \ref{fig:ma_no_and_yes_instance}) to a $\umapp_{yes}$ instance (Fig. \ref{fig:ma_unique_instance}) with inverse polynomial probability, and a $\tmapp_{no}$ instance to a $\umapp_{no}$ instance (left panel in Fig. \ref{fig:ma_no_and_yes_instance}) with probability 1. We divide the potential witnesses into three groups based on their probability of acceptance: 
\begin{align}
Y_{no}=\{y|\ |y|=l\ and\ \Pr(V(x,y)\ accepts\ in\ t\ steps)\leq p_1 \} \notag \\
Y_{gap}=\{y|\ |y|=l\ and\ \Pr(V(x,y)\ accepts\ in\ t\ steps)\in  (p_1,p_2)\} \notag \\
Y_{yes}=\{y|\ |y|=l\ and\ \Pr(V(x,y)\ accepts\ in\ t\ steps) \geq p_2\} \label{eq:y_yes_gap_no}
\end{align}

Let us look at the $R$-restriction of $V$, $V_R$, where $R$ is a random set and each element in $[2^l]$ is taken with some probability $p$.
We denote it by $I'=\langle V_R,x,p_1,p_2,l,t+t'\rangle$, where $t'$ is the time taken for the machine $V_R$ to decide membership in $R$. Define $Y'_{yes},Y'_{gap},Y'_{no}$ for $I'$, as was done in Equation \ref{eq:y_yes_gap_no}.
For every $y$ of length $l$, denote by $f(y)=\Pr(V(x,y)\ accepts\ in\ t\ steps)$, and $f'(y)=\Pr(V'(x,y)\ accepts\ in\ t+t'\ steps)$.
\begin{observation}
\label{ob:ftag}
\begin{equation*}
f'(y) = \begin{cases}
         0 & \text{if } y\notin R\\
         f(y) & \text{if }  y\in R
        \end{cases}
\end{equation*}
Therefore, $Y'_{yes}=Y_{yes}\cap R$ and $Y'_{gap}=Y_{gap}\cap R$.
\end{observation}

Using the same method as in the $\NP$ case fails, as we show next.

\subsection{Problems with the First Component}
\label{sec:problems_with_first_component}
\begin{figure}[htp]
\centering
\scalebox{0.5}{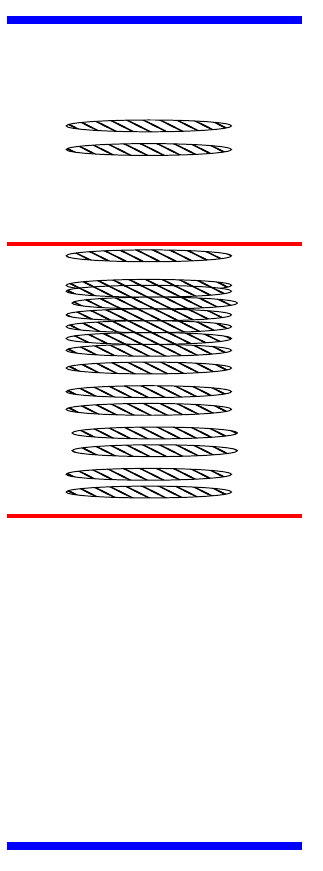} 
\caption{A $\tmapp$ instance for which the first component fails to work: it has numerous witnesses with probability inside the ``gap-interval'' and very few in the ``yes-interval''. See the discussion in Section \ref{sec:problems_with_first_component}.}
\label{fig:ma_problematic}
\end{figure}

We present an instance that shows that implementing component 1 in the probabilistic case fails. The example is a $I^{problematic}=\langle V^{problematic},x,p_1,p_2,l,t\rangle\in \tmapp_{yes}$ instance that can be seen in Fig.\ref{fig:ma_problematic} with the property that $|Y_{yes}|=2, |Y_{gap}|=2^l-2$ and $|Y_{no}|=0$. 

Because the size of the set $Y_{gap}$ is exponentially bigger than that of the set $Y_{yes}$, we cannot ``filter'' one element from $Y_{yes}$ and none from $Y_{gap}$ with non-negligible probability: For example, suppose we pick the size of $R$ based on the set $Y_{yes}$, so each element is chosen with probability $1/2$. With probability $\Omega(1)$, exactly one element will be chosen from $Y_{yes}$, but about half of the elements of $Y_{gap}$ will also be chosen. Therefore, the second property of a $\umapp_{yes}$ instance fails to hold. If we pick elements in $R$ by the size of $Y_{gap}$, so that each element is picked with probability $\frac {1}{2^l -2 }$, then with probability $(1-\frac{1}{2^l-2})^2$ (which is exponentially close to one), no element will be picked from $Y_{yes}$; therefore, the first property of a $\umapp_{yes}$ instance fails to hold. 

\subsection{The Fourth Component}
We first define the notion of a lightweight-gap:
\begin{definition}[``lightweight-gap'' instance]
 An instance $I=\langle  V,x,p_1,p_2,l,t\rangle $ is a ``lightweight-gap'' $\tmapp_{yes}$ instance if it is a $\tmapp_{yes}$ instance, and $|Y_{gap}| < 3 |Y_{yes}|$.  
 \label{def:lightweight_gap_instance}
\end{definition}

Lemma~\ref{le:lwgap_works} explains how lightweight-gap instances are not prone to the problem that was shown in Section \ref{sec:problems_with_first_component}. But first we will see how to create a very simple transformation that takes a general $\tmapp_{yes}$ instance to a ``lightweight-gap'' $\tmapp_{yes}$ instance:
\begin{lemma}
\label{le:transforming_to_lightweight_gap}
 Let $\hat{I}$ be a $\tmapp$ instance. There exists an efficient randomized transformation that maps $I$ to several instances $I_1,...,I_{l-2}$ with the following properties:
\begin{itemize}
 \item If $\hat{I}\in \tmapp_{yes}$ then $\exists k\ s.t.\ I_k$ is a ``lightweight-gap'' $\tmapp_{yes}$ instance.
 \item If $\hat{I}\in \tmapp_{no}$ then $\forall k\ I_k \in \tmapp_{no}$ instance.
\end{itemize}
\end{lemma}
\begin{proof}
 The transformation is as follows. We map $\hat{I}=\langle \hat{V},x,p_1,p_2,l,t\rangle$ to $I=\langle V,x,\frac{1}{l},1-\frac{1}{l},l,t\rangle$ so that:
 \begin{equation}
     \hat{I}\in \tmapp_{yes} \Rightarrow I \in \tmapp_{yes} \text{ and }\hat{I}\in \tmapp_{no} \Rightarrow I \in \tmapp_{no}.
     \label{eq:hatI_implies_I}
 \end{equation}
 This is done by using standard error reduction techniques. Here we crucially use the fact that $p_1$ and $p_2$ are represented in unary, and therefore, $p_1-p_2\geq \frac{1}{poly(n)}$, where $n$ is the input size. 

\begin{observation}
\label{ob:devision}
 Let $I_1=\langle V,x,p_1,p_2,l,t\rangle$ and let $I_2=\langle V,x,q_1,q_2,l,t\rangle$, where $[q_1,q_2]\subseteq [p_1,p_2]$.
\begin{itemize}
 \item $I_1 \in \tmapp_{yes} \Rightarrow I_2\in \tmapp_{yes}$.
\item $I_1 \in \tmapp_{no} \Rightarrow I_2\in \tmapp_{no}$. 
\end{itemize}
\end{observation}
The observation follows immediately from the definitions of $\tmapp$.

The second step of the transformation is as follows: we take the instance $I=\langle V,x,\frac{1}{l},1-\frac{1}{l},l,t\rangle$ and define $l-2$ instances, $I_1,...,I_{l-2}$, where $I_j=\langle V,x,\frac{j}{l},\frac{j+1}{l},l,t\rangle$. By observation \ref{ob:devision}, we know that if $I\in \tmapp_{yes}$ then $\forall k\ I_k\in \tmapp_{yes}$; and that if $I\in \tmapp_{no}$ then $\forall k\ I_k\in \tmapp_{no}$.

We are left to prove that when $I\in \tmapp_{yes}$, one of the $I_k$ is a ``lightweight-gap'' $\tmapp_{yes}$ instance. This will follow from Observation~\ref{ob:existence_of_lightweight_range}:
\begin{observation}[Existence of lightweight range (see also Fig.~\ref{fig:ma_lightweight})]
\label{ob:existence_of_lightweight_range}
 We define $l-1$ ranges: $r_j=[\frac{j}{l},\frac{j+1}{l}),\ 1 \leq j \leq l-2$ and $r_{l-1}=[1-\frac{1}{l},1]$. We define \[Y_j=\{y|\ |y|=l\ and\ \Pr(V(x,y)\ accepts\ in\ t\ steps) \in r_j \}\]
If $I=\langle V,x,\frac{1}{l},1-\frac{1}{l},l,t\rangle\in \tmapp_{yes}$, then there exists a $j\in  \{1,\ldots,l-2\}$  s.t. $|Y_j| < 3|Y_{j+1}| $. 
\end{observation}

\begin{figure}[htp]
\centering
\scalebox{0.5}{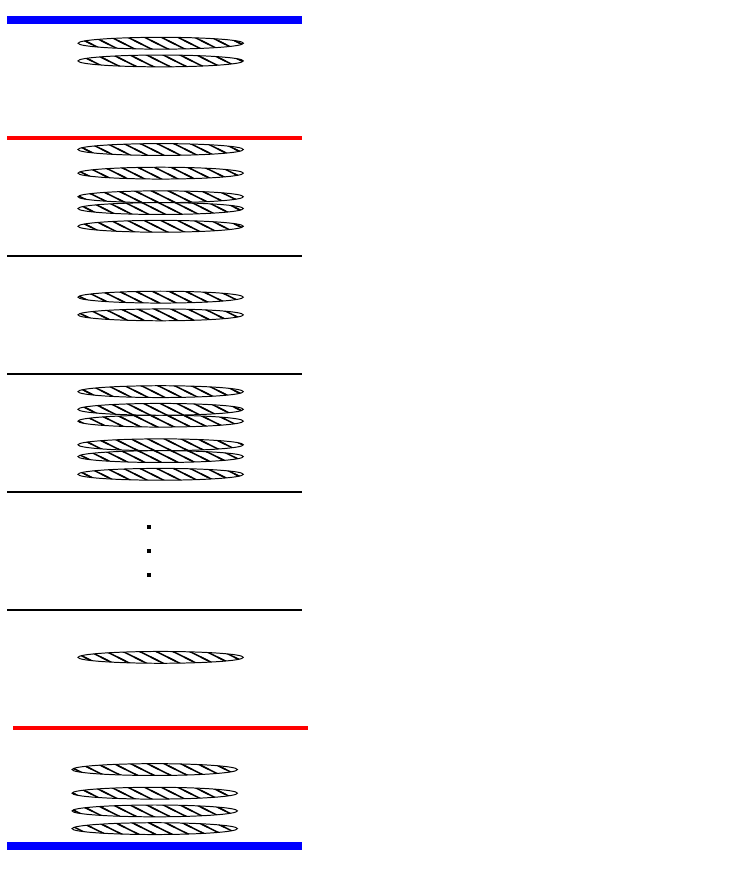}
\caption{A yes-instance with its lightweight range.}
\label{fig:ma_lightweight}
\end{figure}

\begin{proof}
First, notice that $I\in \tmapp_{yes}$ implies $|Y_{l-1}| \geq 1$. Now, assume by contradiction that the inequality does not hold for every $j$, i.e., $|Y_j|\geq 3|Y_{j+1}|$. Then, $|Y_1|\geq 3^{l-2} > 2^l$, where the strict inequality holds for $l\geq 7$, which can be assumed w.l.o.g. The total number of the witnesses is $2^l$, and therefore, $|Y_1|\leq 2^l$. Contradiction.
\end{proof}

To prove Lemma \ref{le:transforming_to_lightweight_gap}, we observe that $|Y_j| < 3|Y_{j+1}|$ for some $j\in \{1,\ldots,l-2\}$, which implies that  $I_j$ is a ``lightweight-gap'' $\tmapp_{yes}$ instance. Observation \ref{ob:existence_of_lightweight_range} asserts that such a $j$ indeed exists, which completes the proof of Lemma~\ref{le:transforming_to_lightweight_gap}.
\end{proof}

The following lemma proves that component 1 works for lightweight-gap $\tmapp_{yes}$ instances:
\begin{lemma}
\label{le:lwgap_works}
Suppose $I=\langle V,x,p_1,p_2,l,t\rangle$ is a lightweight-gap $\tmapp_{yes}$ instance (see Definition~\ref{def:lightweight_gap_instance}). Let $I'=\langle V_R,x,p_1,p_2,l,t+t'\rangle$, where $V_R$ is the R-restriction of $V$, and each element in $R$ is taken with probability $p=\frac{1}{|Y_{gap}|+|Y_{yes}|}$ (see Eq.~\eqref{eq:y_yes_gap_no} for the $Y_{gap}$ and $Y_{yes}$). With probability $\Omega(1)$ (over the choice of R), $I'$  is a $\umapp_{yes}$ instance.
\end{lemma}
\begin{proof}
As was shown for component 1, exactly one witness will be picked from the set $Y_{yes} \cup Y_{gap}$ with probability $\Omega(1)$ . The probability that the instance is from the set $Y_{yes}$ is proportional to the set's size. Therefore, $\Pr(I'\in \umapp_{yes})=\Omega(1) \frac{|Y_{yes}|}{|Y_{gap}|+|Y_{yes}|}\geq \frac{1}{4} \Omega(1)=\Omega(1)$.
\end{proof}

Component 2 works without any change in the probabilistic setting: a constant approximation of the size $|Y_{yes}|$ is sufficient. To adapt component 3 to the current setting, we need a simple variant of Lemma \ref{le:vv}:
\begin{lemma}
\label{le:alternative_vv}
Let $S\subseteq \{0,1\}^l$ of size $b$, and $j\in \NN$ such that $2^j \leq b < 2^{j+1}$, $S_1\subset S$ of size $a$, and $S_2=S \setminus S_1$. 
Let h be a random function sampled from a pairwise independent hash function family  $\mathbb{H}_{l,j+2}$. Then, 
\begin{equation*}
 \Pr\left(|h^{-1}(0)\cap S_1| = 1 \wedge |h^{-1}(0)\cap S_2| = 0\right) \geq \frac{a}{8b}.
\end{equation*}
\end{lemma}
The proof is given in Appendix \ref{sec:proofs}. We apply Lemma \ref{le:alternative_vv} to our construction by setting $S_1=Y_{yes},\ S_2=Y_{gap},\ S=S_1\cup S_2$.

We now have all the tools needed to prove Theorem~\ref{thm:uma_problem_which_is_MA_hard}.

\begin{proofof}{Theorem~\ref{thm:uma_problem_which_is_MA_hard}}
As mentioned in the beginning of this section, we show that $\tmapp\leq_r\umapp$. The randomized reduction is given in Algorithm~\ref{alg:tmapp_to_umapp}. 

\begin{algorithm}[H]
\DontPrintSemicolon
\label{alg:tmapp_to_umapp}
  \SetAlgoLined
   \KwIn{ $\hat I=\langle \hat V,x,p_1,p_2,l,t\rangle$}
    Construct $I=\langle V,x,\frac{1}{l},1-\frac{1}{l},l,t\rangle$ so that $\hat{I}\in \tmapp_{yes} \Rightarrow I \in \tmapp_{yes}$ and $\hat{I}\in \tmapp_{no} \Rightarrow I \in \tmapp_{no}$ (see Eq~\eqref{eq:hatI_implies_I}) \; 
    Sample $k$ uniformly at random from $\{1,\ldots,l-2\}$\quad ($k$ represents the guess of the lightweight range) \;
    Sample $j$ uniformly at random from $\{0,\ldots,l\}$ \;
    Sample $h$ uniformly at random from a family of efficient pairwise independent hash functions $\mathbb{H}_{l,j+2}$, and let $R_j=h^{-1}(0)$\;
    \KwOut{$I_{k,j}=\langle V_{R_j},x,\frac{k}{l},\frac{k+1}{l},l,t+T_{l,j+2}\rangle$ }
\caption{Randomized reduction from $\tmapp$ to $\umapp$.}
\end{algorithm}

It is obvious that the algorithm runs in (randomized) polynomial time. For the soundness, we have that $\forall k,b\ I\in \tmapp_{no}\Rightarrow I_{k,j}\in \tmapp_{no}$, by using observation \ref{ob:devision} and observation \ref{ob:ftag}. Finally, let us analyze the completeness of the reduction. By Eq.~\eqref{eq:hatI_implies_I}, we have $I\in \tmapp_{yes}$. According to lemma \ref{le:transforming_to_lightweight_gap}, for some $k$, $I_k$ is a ``lightweight-gap'' $\tmapp_{yes}$ instance, so this $k$ is chosen with probability $\frac{1}{l-2}$. Define $Y^k_{yes},Y^k_{gap}$ for $I_k$ in a similar manner to Equation (\ref{eq:y_yes_gap_no}). According to Lemma \ref{le:alternative_vv}, for $k$ such that with $S_1=Y^k_{yes},\ S_2=Y^k_{gap},\ S=S_1\cup S_2$, we have:
\begin{equation}
\Pr(|h^{-1}(0)\cap Y_{yes}^k|=1 \wedge |h^{-1}(0)\cap Y_{gap}^k|=0) \geq \frac{a}{8b}=\frac{|Y_{yes}^k|}{8(|Y_{yes}^k|+|Y_{gap}^k|)}\geq \frac{1}{32},    
\label{eq:one_element_from_yes}
\end{equation}
where in the last inequality we used the fact that $I_k$ is a lightweight-gap yes instance, and therefore, $|Y_{gap}^k|< 3|Y_{yes}^k| $. If the event in Eq.~\eqref{eq:one_element_from_yes} holds, we have exactly one witness that gets accepted with probability at least $\frac{k+1}{l}$ and no witnesses that get accepted in the gap region $[\frac{k}{l},\frac{k+1}{l})$, which makes it a $\umapp_{yes}$ instance.
Overall, if $\hat I \in\tmapp_{yes}$,  it is mapped to a $\umapp_{yes}$ instance with probability at least $\frac{1}{32l}$.
\end{proofof}
\section{Valiant-Vazirani Extended to the Class \QCMA}
\label{sec:vv_to_qcma}

The proof of Theorem \ref{thm:uqcma_problem_which_is_QCMA_hard} is almost identical to that for the $\MA$ case (Theorem~\ref{thm:uma_problem_which_is_MA_hard}) but with some minor adaptations discussed below.
By exactly the same argument as in Corollary~\ref{cor:np_subseteq_rp_up}, Theorem~\ref{thm:uqcma_problem_which_is_QCMA_hard} implies the following:
\begin{corollary}
$\QCMA \subseteq \RP^\UQCMA$. 
\end{corollary}

We define the \QCMA\ analogue of \tmapp\ and \umapp\ to be:
\begin{definition}[\tqcmapp]
$\tqcmapp=(L_{yes},L_{no})$. The words in \tqcmapp\ are tuples, $\langle U,p_1,p_2 \rangle$, where $U$ is a description of a quantum circuit with input size denoted by $l$, and $p_1,p_2$ are given in unary s.t.:
\begin{enumerate}
\item $\langle U,p_1,p_2 \rangle \in L_{yes}$ if there exists a string $y$ of length l, s.t. $\Pr(U\ accepts\ \ket{y})\geq p_2$.
\item  $\langle U,p_1,p_2 \rangle \in L_{no}$ if for all strings $y$ of length l, $\Pr(U\ accepts\ \ket{y})\leq p_1$.
\end{enumerate}
\label{def:tqcmapp}
\end{definition}

\begin{definition}[\uqcmapp]
$\uqcmapp=(L_{yes},L_{no})$. The words in \uqcmapp\ are tuples, $\langle U,p_1,p_2 \rangle$, where $U$ is a description of a quantum circuit with input size denoted by $l$, and $p_1,p_2$ are given in unary s.t.:
\begin{enumerate}
\item $\langle U,p_1,p_2 \rangle \in L_{yes}$ if there exists a string $y$ of length l, s.t. $\Pr(U\ accepts\ \ket{y})\geq p_2$ and $\forall y'\neq y\ \Pr(U\ accepts\ \ket{y})\leq p_1$.
\item  $\langle U,p_1,p_2 \rangle \in L_{no}$ if for all strings $y$ of length l $\Pr(U\ accepts\ \ket{y})\leq p_1$.
\end{enumerate}
\end{definition}
$\tqcmapp$ is (trivially) $\QCMAC$, and similarly, $\uqcmapp$ is $\UQCMAC$. 

Equipped with the definitions above and all of the steps done in the proof of Theorem~\ref{thm:uma_problem_which_is_MA_hard}, we are ready to that $\tqcmapp\leq_r \uqcmapp$. 
This reduction is shown in Algorithm~\ref{alg:tqcmapp_to_uqcmapp}. 


\begin{algorithm}[H]
\DontPrintSemicolon
\label{alg:tqcmapp_to_uqcmapp}
\SetAlgoLined
    \KwIn{ $\hat I=\langle \hat U,p_1,p_2\rangle$}
    Construct $I=\langle U,\frac{1}{l},1-\frac{1}{l}\rangle$ so that $\hat{I}\in \tqcmapp_{yes} \Rightarrow I \in \tqcmapp_{yes}$ and $\hat{I}\in \tqcmapp_{no} \Rightarrow I \in \tqcmapp_{no}$ \; 
    Sample $k$ uniformly at random from $\{1,\ldots,l-2\}$\quad ($k$ represents the guess of the lightweight range) \;
    Sample $j$ uniformly at random from $\{0,\ldots,l\}$ \;
    Sample $h$ uniformly at random from a family of efficient pairwise independent hash functions $\mathbb{H}_{l,j+2}$, and let $R_j=h^{-1}(0)$\;
    \KwOut{$I_{k,j}=\langle U_{R_j},\frac{k}{l},\frac{k+1}{l}\rangle$ }
\caption{Randomized reduction from $\tqcmapp$ to $\uqcmapp$.}
\end{algorithm}

In the output of the algorithm, we define $U_{R_j}$ in an approach analogous to that used for $\MA$. Specifically, we begin by testing whether $y\in R_j$. If it is not, it rejects, and otherwise, it applies $U$.  Of course, this can be done efficiently with a quantum circuit, since we use an efficient hash functions.

Soundness and Completeness follow from the same arguments used in the \MA\ case. This ends the proof of Theorem \ref{thm:uqcma_problem_which_is_QCMA_hard}.

\section{Robustness of \UQMA}
\label{sec:robustness}

\subsection{Discussion about QMA and the Marriott-Watrous Formalism}
\label{sec:marriott-watrous}
In this section, we discuss the robustness of our definition of unique $\QMA$ and prove Theorem \ref{thm:onedulh_is_uqmac}.

From Definition \ref{QMAdef}, we see that for a given \QMA\ verification scheme and a state $\ket{\psi}$, its probability of acceptance is:
\begin{equation*}
\Pr( \text{verifier accepts } \ket{\psi})=\lVert \Pi_1 U_x(I\otimes\ket{0^m})\ket{\psi} \rVert^2 
\end{equation*}
A useful operator in this context, as defined in \cite{MW05}, is the following
\begin{equation} \label{Qdef}
Q=(I_m\otimes \bra{0^m}) U^\dagger \Pi_1 U(I\otimes\ket{0^m}).
\end{equation}
Note that
\begin{equation}
\label{eq:acceptance_probability}
\Pr( \text{verifier accepts }\ket{\psi})=\bra{\psi}Q\ket{\psi}.
\end{equation}
            As $Q$ is Hermitian, there is a basis of orthonormal eigenvectors $\{\ket{\psi_i}\}_{i=1}^{2^l}$ for which $Q=\sum_i \lambda_i \ketbra{\psi_i}{\psi_i}$, where $\lambda_i(Q) \geq \lambda_{i+1}(Q)$ are the eigenvalues of $Q$. Note that by knowing the eigenvectors and eigenvalues of $Q$, we can find out the acceptance probability of every witness in a simple way
\begin{align}
 \label{eq:convex_combination}
 \bra{\psi} Q \ket{\psi} &= \sum_{i,j} a_i^*a_j  \bra{\psi_i} Q \ket{\psi_j}\\
&=\sum_{i,j} a_i^*a_j \lambda_j  \braket{\psi_i}{\psi_j} =\sum_i |a_i|^2 \lambda_i, \nonumber
\end{align}
where $a_i=\braket{\psi_i}{\psi}$.

Let us consider another possible definition of the class $\UQMA$.

\begin{definition}[Alternative definition for \UQMA]
\label{def:uqma_watrous}
A promise problem $L=(L_{yes},L_{no}) \in \UQMA$ if there exists a polynomial quantum circuit $U_x$ that can be computed in $poly(|x|)$ time, having $l(x)$ qubits as input and requiring $m(x)$ ancilla qubits initialized to $\ket{0^m}$, s.t.
\begin{enumerate}

 \item $x\in L_{yes} \Rightarrow \lambda_1(Q) \geq 2/3 \text{ and } \lambda_2(Q) \leq 1/3$.
\item $x\in L_{no} \Rightarrow \lambda_1(Q) \leq 1/3$.
\end{enumerate}
Where $\lambda_1 \geq \lambda_2 \geq \ldots \lambda_{2^{l(x)}}$ are the eigenvalues of $Q$.
\end{definition}

\begin{lemma}[Equivalence of Definitions \ref{def:uqma} and \ref{def:uqma_watrous}]
A language $L=(L_{yes},L_{no})\in \UQMA$ according to Definition \ref{def:uqma} $ \Longleftrightarrow L\in \UQMA$ according to Definition \ref{def:uqma_watrous}.
\label{lem:uqma_definitions_are_equivalent}
\end{lemma}
\begin{proof}
We start by proving that, given a $I\in L_{yes}$ according to Definition \ref{def:uqma}, it is also in $L_{yes}$ according to Definition \ref{def:uqma_watrous}. We know from Definition \ref{def:uqma} that there is state $\ket{\psi}$ that is accepted with probability of at least $\frac{2}{3}$. According to Eq. (\ref{eq:acceptance_probability}), the acceptance probability of $\ket{\psi}$ is $\bra{\psi}Q\ket{\psi}= p\geq 2/3$. From Eq. (\ref{eq:convex_combination}), in turn, we see that $p$ can be written as a convex combination of the $\lambda$'s. Therefore, $\lambda_1 \geq 2/3$. 

We now prove that $\lambda_2 \leq 1/3$. Denote by $V$ the subspace spanned by the eigenvectors with eigenvalue greater than $1/3$. Note that $\forall \ket{\varphi} \in V \ \bra{\varphi}Q\ket{\varphi} > 1/3$. If $dim(V)\geq 2$, there must exist a $\ket{\varphi}\in V$ orthogonal to $\ket{\psi}$, and therefore, the acceptance probability of $\ket{\varphi}$ is greater than $1/3$, which is in contradiction to the properties of an $L_{yes}$ instance according to definition \ref{def:uqma}.

The other direction is straightforward.
\end{proof}


We now turn to the proof of Theorem \ref{thm:onedulh_is_uqmac}. Let us start with the precise definition of the problem \onedulh:

\begin{definition}
$\onedulh$: We are given a $2$-local Hamiltonian on $n$ $d$-dimensional sites $H = \sum_{j=1}^r H_j$ with $r = \poly(n)$ arranged in a line. Each $H_j$ has a bounded operator norm $|| H_j || \leq \poly(n)$. We are also given two constants, $a$ and $b$, with $b - a \geq 1/\poly(n)$. In ``yes'' instances, the smallest eigenvalue of $H$ is at most $a$, and all the other eigenvalues are above $b$. In ``no'' instances, the smallest eigenvalue is larger than $b$. We need to decide whether a given instance is a "yes" or a "no" instance. 
\end{definition}

\begin{proofof}{Theorem~\ref{thm:onedulh_is_uqmac}} 
\label{pf:onedulh_is_uqmac}
From the following verification procedure, it is evident that the problem is in $\UQMA$. As a proof, we expect the unique ground state of $H$. Given a witness $\ket{\psi}$, we use the phase estimation algorithm (see e.g., Ref. \cite{WZ06}) to determine its energy, i.e., $\bra{\psi}H\ket{\psi}$, within inverse polynomial accuracy $\delta$. The algorithm has the property that if $\ket{\psi}$ is an eigenstate of $H$, then the output will be the eigenvalue (up to accuracy $\delta$) with exponentially high probability. 

If the output of the phase estimation is smaller than $a + \delta$, we accept; otherwise we reject. It is clear that in ``yes'' instances, there is exactly one witness that is accepted with probability exponentially close to one (the ground state of $H$), while any state orthogonal to it is accepted only with an exponentially small probability (which is the probability that the phase estimation does not give the correct answer).  

The hardness of the problem for $\UQMA$ is a simple application of the construction of \cite{AGIK07}, which presents a reduction from any problem in $\QMA$ to a $\onedlh$ with $d=13$\footnote{Aharonov et al.~\cite{AGIK07} originally claimed to prove this result with $d=12$, but Hallgren et al.~\cite{HNN13} note that due to an error, the dimension should be increased by $1$, i.e., their result holds for $d=13$.}. The details of the construction are not important here. We only note that the low-lying eigenvectors of the Hamiltonian considered are well approximated, within an inverse polynomial, to a class of states parametrized by all possible proofs --- called history states --- with the property that two orthogonal proofs give raise to two orthogonal history states. Moreover, the probability of acceptance of a given proof is imprinted in the energy of the associated history state, which again holds up to inverse polynomial accuracy. It is then clear that a problem in $\UQMA$ will give rise to valid instance of $\onedulh$, since in ``yes'' instances of the problem (which is the only case we must analyze), the second eigenvalue of the Hamiltonian, which is well approximated by the energy of the history state associated with the witness that has the \textit{second} highest probability of acceptance, will be separated from the ground state energy by a constant factor.
\end{proofof}

\subsection{Yet Another New Class and Its Equivalence To \texorpdfstring{$\UQMA$}{\textsf{UQMA}}}

One might define a class that is similar to $\QMA$, with the added promise of the gap in its acceptance probability.

\begin{definition}[\textsf{Poly-Gapped} $\QMA$ (\PGQMA)]
A promise problem $L=(L_{yes},L_{no}) \in \PGQMA$ if there exist $\delta(|x|)=1/poly(|x|)$, and a polynomial quantum circuit $U_x$ that can be computed in $poly(|x|)$ time, having $l(x)$ qubits as input and requiring $m(x)$ ancilla qubits initialized to $\ket{0^m}$, s.t.
\begin{enumerate}
 \item $x\in L_{yes} \Rightarrow \lambda_1(Q) \geq 2/3 \text{ and } (\lambda_1(Q) - \lambda_2(Q)) \geq \delta(|x|)$.
\item $x\in L_{no} \Rightarrow \lambda_1(Q) \leq 1/3 \text{ and } (\lambda_1(Q) - \lambda_2(Q)) \geq {\delta(|x|)}$.
\end{enumerate}
Here, $\lambda_1(Q) \geq \lambda_2(Q) \geq \ldots \lambda_{2^{l(x)}}(Q)$ are the eigenvalues of the operator $Q$, defined in Eq. (\ref{Qdef}).
\end{definition}

The above definition is motivated by the $\lh$ problem, with the additional promise that the spectral gap of the Hamiltonian is inverse polynomial. Its one-dimensional version is defined as follows. 

\begin{definition}
$\onedpglh$: We are given a $2$-local Hamiltonian on $n$ $d$-dimensional sites $H = \sum_{j=1}^r H_j$ with $r = \poly(n)$ arranged in a line. Each $H_j$ has a bounded operator norm $|| H_j || \leq \poly(n)$. We are also given three constants, $a$, $b$ and $\Delta$, with $b - a, \Delta \geq 1/\poly(n)$. We have the promise that the spectral gap of $H$ is at least $\Delta$. In ``yes'' instances, the smallest eigenvalue of $H$ is at most $a$. In ``no'' instances, the smallest eigenvalue is at least $b$.  We need to decide whether a given instance is a "yes" or a "no" instance.
\end{definition}

As in the unique case, we can show,
\begin{theorem}
$\onedpglh$ is \PGQMAC.
\label{thm:onedpglh_is_PGQMAC}
\end{theorem}
The proof is completely analogous to the reasoning we provided for Theorem \ref{thm:onedulh_is_uqmac}.

\begin{lemma}
$\UQMA \subseteq \PGQMA$. 
$\PGQMA \subseteq \RP^\UQMA$.
\label{le:pgqma_equals_uqma_randomized}
\end{lemma}

\begin{proof}
We first show that $\UQMA \subseteq \PGQMA$. This inclusion is not immediate because of the following reason:
If $I \in L_{no} \in \UQMA$, then we know that $\lambda_1(Q) \leq 1/3$, but it is not guaranteed that $(\lambda_1(Q) - \lambda_2(Q)) \geq {\delta(x)}$.

To resolve this issue, we use the amplification property of \QMA\  and change the ``no''-probability to be $1/3 - \delta$ instead of $1/3$ so that we have $\lambda_1(Q) \leq 1/3 - \delta$. Then, by a simple construction that we explain below, we add a single state that is accepted with probability $1/3$, having $\lambda_1(Q) = 1/3$ and $\lambda_2(Q) \leq 1/3 - \delta$, which provides the necessary gap.

Adding the $1/3$-eigenvalue is done by changing the circuit: we append another qubit to the input qubits, and measure it in the beginning of the circuit. If the outcome is 0, then we proceed as before; If it is 1, we measure all other $l$ input qubits in the computational basis. If all of them are 1, we accept with probability 1/3, and reject otherwise. A simple calculation shows that the action of such a procedure is exactly what we want: it adds a single $1/3$-eigenvalue (for the state $\ket{1}\otimes \ket{1^l}$), and $2^l - 1$ 0-eigenvalues (for states of the form $\ket{1}\otimes \ket{z}$ for $z\neq 1^l$), which do not concern us. The other eigenvalues remain the same: If the state $\ket{\psi}$ had an eigenvalue $\lambda$ in the original circuit, then the state $\ket{0}\otimes \ket{\psi}$ has the same eigenvalue in the modified circuit. 

We now show that $\PGQMA \subseteq \RP^\UQMA$. Again, this is not immediate, but this time due to the completeness: for $I \in L_{yes} \in \PGQMA$, we know that $\lambda_1(Q) \geq 2/3$, but we do not know whether $\lambda_2(Q)$ is below the ``no''-probability.	For this we use the fact that $\UQMA_{1/3,2/3} = \UQMA_{a,b}$, where $(b-a)\geq 1/poly$. We know that for a $I\in L_{yes}$ there exists a $j\in \mathbb N$, $j\leq \frac{1}{\delta(|x|)}$ for which $\lambda_1(Q) \geq 2/3+(j+1)\frac{\delta}{2}$ and $\lambda_2(Q) \leq 2/3+j\frac{\delta}{2}$. Thus, the $\RP$ algorithm guesses $j\leq \frac{1}{\delta(|x|)}$ and makes a query to the $\UQMA_{2/3 + j\delta /2 ,2/3 + (j+1)\delta/2}$. At least for one $j$, the answer to the query will be ``yes'', in which case we accept. 
\end{proof}
We are now ready to prove Theorem~\ref{thm:onedplh_is_qcma_hard}.

\begin{proofof}{Theorem~\ref{thm:onedplh_is_qcma_hard}}
\label{pf:onedplh_is_qcma_hard}
We use the same proof as in the second part of the proof in Lemma~\ref{le:pgqma_equals_uqma_randomized}, to get the reduction $\onedpglh \leq_r \onedulh$. We already know that the $\onedulh$ problem is $\QCMA$-hard under randomized reductions by Theorem~\ref{thm:uqcma_problem_which_is_QCMA_hard}. Using the transitivity of randomized reductions (see Observation~\ref{ob:proerties_of_randomized_reductions}.\ref{it:random_reduciton_transitive}), we can combine these two reductions and conclude that $\onedpglh$ is $\QCMA$-hard, as required.
\end{proofof}

\section{The \QMA\ Case}
\label{sec:qma_impossibility_results}

\subsection{Random Projections Fail to Create Inverse Polynomial Gap}

As mentioned earlier, we divided the proof of the Valiant-Vazirani Theorem into three components. Component 1 solves the problem in the simple case where the number of the accepting witnesses is known; Component 2 improves it by observing that the size of the set can only be approximated, without considerably affecting the probability of acceptance; Finally, Component 3 shows that we may achieve the same results by using a pairwise independent hash function instead of a random function, thus rendering the reduction efficient.  

In this section, we show that even in the case where the number of solutions is known, as in component 1, we cannot --- at least in the most direct approach  --- create a transformation that maps it to a ``unique instance''. The main difficulty in the \QMA\ case is that we do not know in which basis to operate. Notice that if there exists a description (that Merlin can supply) of how to efficiently transform a standard basis state to one of the states that is accepted with probability greater than $\frac{2}{3}$, then the problem is in $\QCMA$. 

Let us define a possible quantum analogue of an $R$-restriction. A natural generalization --- rather than restricting to witnesses that belong to some set $R$ --- is to project onto some subspace $R$; We call this procedure a quantum $R$-restriction. As we did in the discussion of component 1, we will not consider the efficiency of implementing the restriction. A diagram of a general circuit and its $R$-restriction are given in Figure \ref{figdiagram}.

\begin{figure}
\centerline{
  \Qcircuit @C=1em @R=1em {
& \qw & \multigate{5}{U} & \meter & &\\
\lstick{\ket{\psi}} &\vdots  &  & & &\\
& \qw & \ghost{U} & \qw & &\\
\lstick{\ket{0}} & \qw & \ghost{U} & \qw& &\\
& \vdots & & & &\\
\lstick{\ket{0}} &  \qw & \ghost{U} & \qw & &
} \quad \quad
\Qcircuit @C=1em @R=1em {
& \qw & \multigate{5}{U} & \meter & \multigate{5}{\Pi_R} & \\
\lstick{\ket{\psi}} &\vdots  &  & & &\\
& \qw & \ghost{U} & \qw & \ghost{\Pi_R}&\\
\lstick{\ket{0}} & \qw & \ghost{U} & \qw&\ghost{\Pi_R} &\\
& \vdots & & & &\\
\lstick{\ket{0}} &  \qw & \ghost{U} & \qw & \ghost{\Pi_R}&
}
}
\caption{A quantum R-restriction. On the left: a general description of a \QMA\ verification scheme. On the right: its $R$-restriction, where $\Pi_R$ is the projection on the subspace $R$. The state is accepted only if in both measurements the outcome was 1. }
 \label{figdiagram}
\end{figure}

While the relevant operator for the original verification is $Q=(I_l\otimes \bra{0^m})U^\dagger \Pi_1 U (I_l\otimes \ket{0^m})$, after the $R$-restriction, it is given by $Q_R=(I_l\otimes \bra{0^m})U^\dagger \Pi_1 \Pi_R \Pi_1 U (I_l\otimes \ket{0^m})$, where $\Pi_R$ is a projection onto the subspace $R$. The quantum analogue of component 1 consists of taking the subspace $R$ to be a random subspace of dimension $d$, chosen according to the Haar measure, for some convenient $d$. The next proposition shows that this approach, unfortunately, fails. 

\begin{proposition} \label{failQMA}
For every $\epsilon > 0$ and $d \in \mathbb N$, with probability larger than $1 - \epsilon$, applying the quantum random $R$-restriction with dimension $d$, to Example \ref{ex:qma_2_accepting_states} creates an instance with a gap smaller than $\epsilon^{-1}2^{-l/2 + 2}$. 
\end{proposition}
\begin{proof}
As the verification circuit already rejects any state in the orthogonal complement of the two-dimensional subspace $V$, it is clear that we only have to analyze the gap created on states in $V$. 

A rank $d$ random projector can be written as $UP_dU^{\cal y}$, where $U$ is a unitary drawn from the Haar measure and $P_d := \sum_{j=1}^d \ket{j}\bra{j}$. Let $m_V(U, d) := \max_{\ket{\psi} \in V} \bra{\psi}U P_d U^{\cal y}\ket{\psi} - \bra{\psi^{\bot}}U P_d U^{\cal y}\ket{\psi^{\bot}}$, where $\ket{\psi^{\bot}}$ is the --- up to a phase --- unique orthogonal vector to $\ket{\psi}$ in $V$. We consider the following quantity, which gives the expectation value of the gap created by applying the random $R$-projection defined by $U P_d U^{\cal y}$:
\begin{equation} \label{intaverage}
\mathbb{E}_{U \sim \text{Haar}}(m_V(U, d)) = \int_{U(2^{l})} dU m_V(U, d),
\end{equation}
where the integral is taken over the Haar measure of the unitary group $U(2^l)$.

Let $\{ \ket{0}, \ket{1} \}$ be a basis for $V$. Note that $m_V(U, d)$ is given by the difference of the maximum $\lambda_{\max}$ and minimum $\lambda_{\min}$ eigenvalues of the following matrix 
\begin{equation*}
V_{U,k} := \left( 
\begin{array}{cc}
\bra{0}U P_d U^{\cal y}\ket{0} & \bra{0}U P_d U^{\cal y}\ket{1} \\
\bra{1}U P_d U^{\cal y}\ket{0} & \bra{1}U P_d U^{\cal y}\ket{1}
\end{array}
\right)
\end{equation*}
By Gershgorin disc Theorem (\cite[p. 244]{Bha97} ), we find
\begin{equation*}
|\lambda_{\max}(V_{U,k}) - \lambda_{\min}(V_{U, k})| \leq |\bra{0}U P_d U^{\cal y}\ket{0} - \bra{1}U P_d U^{\cal y}\ket{1}| + 2 |\bra{0}U P_d U^{\cal y}\ket{1}|,
\end{equation*}
from which follows that
\begin{equation*}
\int_{U(2^{l})} dU m_V(U, d) \leq  \int_{U(2^{l})} dU |\bra{0}U P_d U^{\cal y}\ket{0} - \bra{1}U P_d U^{\cal y}\ket{1}| + 2  \int_{U(2^{l})} dU |\bra{0}U P_d U^{\cal y}\ket{1}|.
\end{equation*}
Applying Lemma \ref{ineqqmau} to each of the two terms in the R.H.S. of the equation above,
\begin{equation*}
\int_{U(2^{l})} dU m_V(U, d) \leq \sqrt{\frac{2k(2^l - k)}{(2^l+1)2^l(2^l-1)}} + 2 \sqrt{\frac{k(2^l - k)}{(2^l+1)2^l(2^l-1)}} \leq 2^{-l/2 + 2},
\end{equation*}
for any $1 \leq k \leq 2^l$. To complete the proof, note that by Markov's inequality,
\begin{equation*}
\int_{U : m_V(U, d) \geq \lambda} dU  \leq 2^{-l/2 + 2}/\lambda,
\end{equation*}
for every $\lambda > 0$. Setting $\lambda = 2^{-l/2 + 2}/\epsilon$, we find that with probability 
\begin{equation*}
\int_{U : m_V(U, d) < \lambda} dU = 1 - \int_{U : m_V(U, d) \geq \lambda} dU \geq 1 - \epsilon,
\end{equation*}
$m_V(U, d)$ is smaller than $2^{-l/2 + 2}/\epsilon$.
\end{proof}

\begin{lemma} \label{ineqqmau}
For any traceless operator $X \in {\cal B}(\mathbb{C}^N)$,
\begin{equation}
\int_{U(N)} dU |\trace(UP_kU^{\cal y}X)| \leq \sqrt{\frac{k(k - K)\trace(X^{\cal y}X)}{(N+1)N(N-1)}},
\end{equation}
where $P_k := \sum_{j=1}^k\ket{j}\bra{j}$.
\end{lemma}

\begin{proof}
From the convexity of the square function,
\begin{equation*}
\left( \int_{U(N)} dU |\trace(UP_kU^{\cal y}X)| \right)^2 \leq \int_{U(N)} dU |\trace(UP_kU^{\cal y}X)|^2. 
\end{equation*}
To compute the R.H.S. of the equation above, we first note that
\begin{align} \label{eq2}
\int_{U(N)} dU |\trace(UP_kU^{\cal y}X)|^2 &=& \int_{U(N)} dU \trace(U^{\otimes 2} P_k^{\otimes 2} (U^{\cal y})^{\otimes 2} X \otimes X^{\cal y}) \nonumber \\ 
&=& \trace( \left(\int_{U(N)} dU  U^{\otimes 2} P_k^{\otimes 2} (U^{\cal y})^{\otimes 2} \right) X \otimes X^{\cal y}).
\end{align}
By Schur's Lemma \cite{FH04} (see also~\cite[pp. 417--418]{Wat18},
\begin{align*}
\int_{U(N)} dU  U^{\otimes 2} P_k^{\otimes 2} (U^{\cal y})^{\otimes 2} &=&  \trace\left(P_k^{\otimes 2}(\id - \text{SWAP})\right) \frac{\id - \text{SWAP}}{N(N-1)} \\ &+& \trace\left(P_k^{\otimes 2}(\id + \text{SWAP})\right) \frac{\id + \text{SWAP}}{N(N+1)} \nonumber \\ &=& \frac{k(k-1)}{N(N-1)}(\id - \text{SWAP}) + \frac{k(k+1)}{N(N+1)}(\id + \text{SWAP}),  
\end{align*}
where $\text{SWAP}$ is the swap operator, and we used that $\trace(\text{SWAP}(P_k \otimes P_k)) = \trace(P_k^2) = \trace(P_k) = k$. Then, from Eq. (\ref{eq2}),
\begin{equation*}
\int_{U(D)} dU \trace(UP_kU^{\cal y}X)^2 =  \trace(X^{\cal y}X)\left(\frac{k(k+1)}{N(N+1)} -  \frac{k(k-1)}{N(N-1)}\right),
\end{equation*}
from which the lemma easily follows.
\end{proof}

\subsection{Using a Many-Outcome Measurement} 

In the previous section we tried to solve example \ref{ex:qma_2_accepting_states} by applying the most natural idea that comes to mind: do a random 2-outcome measurement, and see if one state can ``pass'' the projection with an amount that is not negligible, compared to the other state on the subspace. We found out that such a procedure fails. In this section, we analyze the use a many-outcome measurement. We begin by applying a measurement in a random basis (or, to put it differently, by applying a random unitary according to the Haar measure, and then measuring in the standard basis). This, of course, cannot be done efficiently, but we will deal with it later. 

Radhakrishnan et al. \cite{RRS05} have shown,
\begin{theorem} 
\cite{RRS05} Let $\ket{\psi_1},\ket{\psi_2}$ be two orthogonal quantum states in $\mathbb{C}^N$. Then,
 \[\mathbb{E}_{\hat{M}}\left(\left\|\hat{M}(\ket{\psi_1})-\hat{M}(\ket{\psi_2})\right\|_{1} \right) = \Omega(1) \]
 where $\hat{M}$ is an orthogonal basis chosen uniformly from the Haar measure.
\end{theorem}
A stronger result was presented in \cite[Theorem 1]{Sen06}, which implies the same kind of result, but instead of the expectation, it asserts that the same conditions hold with all but an exponentially small probability.

Furthermore, Ambainis and Emerson \cite{AE07} have shown that:

\begin{theorem} \label{design}
Let $\ket{\psi_1},\ket{\psi_2}$ be two orthogonal quantum states in $\mathbb{C}^N$. Then,
 
  \[ \left\|\hat{M}(\ket{\psi_1})-\hat{M}(\ket{\psi_2})\right\|_{1} = \Omega(1) \]
where $\hat{M}$ is a POVM with respect to an $\epsilon$-approximate (4, 4)-design.
\end{theorem}

For our purposes, one does not need to understand what an $\epsilon$-approximate (4, 4)-design is, but rather, only that an efficient construction exists that enables us to realize the POVM $\hat{M}$ for any constant $\epsilon$. Notice that this is a constant POVM, and for every two states, the TVD of the distributions is constant. For more details of how one can implement a 4-design, see Theorem 1 of \cite{AE07}. Although the POVM is constant, it achieves the same result as a random object (many-outcome measurements), but in an efficient way, and therefore, we see it as a ``pseudorandom'' object.

So how can we exploit that? Suppose we had the description of the distribution of $\hat{M}(\ket{\psi_1})$ and $\hat{M}(\ket{\psi_2})$. Then we could select a unique witness by accepting only when we measure an outcome $j$ that is associated with the $j$'s for which $\hat{M}(\ket{\psi_1})(j) > \hat{M}(\ket{\psi_2})(j)$. As such, we would get by Theorem \ref{design} that $\ket{\psi_1}$ is accepted with a $\Omega(1)$ probability larger than $\ket{\psi_2}$. Of course, this approach does not lead to the solution of the problem, as the promise of having a description of the distributions is too strong. 

Indeed, although there is a classical description that would enable us to distinguish, with high probability, between the two cases, there is no known general way to achieve that which is in $\BQP$. We note that there is a resemblance between this problem and the \SZKC\ problem given in Ref. \cite{Vad99}, in both problems, it is required to distinguish between two probabilities with some total variation distance.

\section{Acknowledgments}
We wish to thank the anonymous reviewers for their suggestions and careful editing.

This work is part of the QIP-IRC supported by EPSRC 
(GR/S82176/0) as well as the Integrated Project Qubit 
Applications (QAP) supported by the IST directorate as 
Contract Number 015848' and was supported by an EPSRC Postdoctoral 
Fellowship for Theoretical Physics.
Most of this work was done while O.S. was at the Hebrew University, and F.G.S.L.B. was at the Imperial College London. 
\bibliographystyle{alphaabbrurldoieprint.bst}
\bibliography{adiabatic}

\appendix 
\section{Ancillary Proofs}
\label{sec:proofs}
\begin{proof-of-lemma}{\ref{le:vv}}
Let $\{y _1,y _2,...,y _{w} \}$ be the elements of $W$. 
\begin{align}
\Pr(|h^{-1}(0)\bigcap W|=1)& = \Pr(\bigcup_{i=1}^w (h(y _i)=0 \bigcap_{j\neq i} h(y _j)\neq 0)) \label{eq:vv1} \\
&=\sum_{i=1}^w \Pr(h(y _i)=0 \bigcap_{j\neq i} h(y _j)\neq 0) \label{eq:vv2}\\
&= \sum_{i=1}^w \Pr(h(y _i)=0)\Pr(\bigcap_{j\neq i} h(y _j)\neq 0 | h(y _i)=0) \notag \\
&=\sum_{i=1}^w \Pr(h(y _i)=0) (1-\Pr(\bigcup_{j\neq i} h(y _j) = 0|h(y _i)=0)) \notag \\
&\geq \sum_{i=1}^w \Pr(h(y _i)=0) (1-\sum_{j \neq i} \Pr(h(y _j)=0|h(y _i)=0)) \label{eq:vv5} 
\end{align}
Equation \eqref{eq:vv2} follows from the fact that all the events in the union of the r.h.s. in Eq.~\eqref{eq:vv1} are disjoint. Equation \eqref{eq:vv5} follows from the union bound.

Since $h$ is sampled from a pairwise independent hash function set,  $\Pr(h(y _i)=0)=\frac{1}{2^{k+2}}$, and $\Pr(h(y _j)=0|h(y _i)=0)=\frac{1}{2^{k+2}}$. Therefore,
\begin{align}
\Pr(|h^{-1}(0)\bigcap W|=1)\geq  \frac{w}{2^{k+2}} (1-\frac{w-1}{2^{k+2}}) \geq \frac{1}{4}(1-\frac{w}{2^{k+2}})\geq \frac{1}{8},
\end{align}
where the last two inequalities follow from the assumption that $2^k\leq w < 2^{k+1}$.
\end{proof-of-lemma}

\begin{proof-of-lemma}{\ref{le:alternative_vv}}

 The proof is very similar to the proof of Lemma \ref{le:vv} above: Let $y _1,...,y _{a}$ be the elements of $S_1$, and $y _{a+1},...,y _{b}$ the elements of $S_2$. So, 
 \begin{align*}
 &\Pr(|h^{-1}(0)\bigcap S_1| = 1 \wedge |h^{-1}(0)\bigcap S_2| = 0) \\
 &=\Pr(\bigcup_{i=1}^a (h(y _i)=0 \bigcap_{1 \leq i' \leq b, i'\neq i} h(y_{i'})\neq 0)).
 \end{align*}
 The next steps are exactly the same as the ones between Eq.~\eqref{eq:vv1} to ~\eqref{eq:vv5}, so we get:
 \begin{align*}
 \Pr(|h^{-1}(0)\bigcap S_1| = 1 \wedge |h^{-1}(0)\bigcap S_2| = 0) &\geq \sum_{i=1}^a \Pr(h(y _i)=0) (1-\sum_{1 \leq i' \leq b, i' \neq i} \Pr(h(y_{i'})=0|h(y_i)=0))\\
 &=  \frac{a}{2^{j+2}} \left(1-\frac{b-1}{2^{j+2}}\right) \geq \frac{a}{4b} \left(1-\frac{b}{2^{j+2}}\right)\geq \frac{a}{8b}, 
 \end{align*}
 where in the last two inequalities we used $2^j \leq b < 2^{j+1}$.
\end{proof-of-lemma}

\end{document}